\documentclass[12pt,floatfix,aps,amsmath,amssymb,preprintnumbers,nofootinbib, notitlepage, prepint]{revtex4-1}
\pdfoutput=1
\usepackage[]{graphicx}
\graphicspath{{./plots/}}
\usepackage{color}
\usepackage{hyperref}					
\usepackage{mathtools}					
\usepackage{slashed}				
\usepackage{algorithm,algorithmic}	
\usepackage{dcolumn}	% Align table columns on decimal point
\usepackage{bm}		% bold math
\usepackage{bbm}
\usepackage{pxfonts}
\usepackage{subfigure}

\usepackage{hyperref}
\usepackage{url}
	
\usepackage[vcentermath]{youngtab}

\def\al#1{\alpha_{#1}}

\usepackage{bbold}

\newcommand{\Tr}{\textrm{Tr}\,}

\newcommand{\Dslash}{\ensuremath \raisebox{0.025cm}{\slash}\hspace{-0.28cm} D}
\newcommand{\be}{\begin{equation}}
\newcommand{\ee}{\end{equation}}
\newcommand{\bea}{\begin{eqnarray}} % only untightened
\newcommand{\eea}{\end{eqnarray}}

\newcommand{\bmp}{\noindent\begin{minipage}{16cm}}
\newcommand{\emp}{\end{minipage}\vskip 7mm} % 7mm untightened
\def\lsim{\mathrel{\raise.3ex\hbox{$<$\kern-.75em\lower1ex\hbox{$\sim$}}}}
\def\gsim{\mathrel{\raise.3ex\hbox{$>$\kern-.75em\lower1ex\hbox{$\sim$}}}}

\newcommand{\intron}[1]{}%{#1}

\definecolor{rossoCP3}{cmyk}{0,.88,.77,.40}
\definecolor{light_gray}{rgb}{0.8,0.8,0.8}

\newcommand{\ea}[1]{\begin{align} #1 \end{align}}

\long\def\del #1 \enddel { }

\begin{document}

\title{ \Large  \color{rossoCP3} Four-Fermion Limit of Gauge-Yukawa Theories }
\author{Jens Krog$^{\color{rossoCP3}{\spadesuit}}$}
\email{krog@cp3-origins.net} 
\author{Matin  Mojaza$^{\color{rossoCP3}{\vardiamondsuit}}$}
\email{mojaza@nordita.org} 
\author{Francesco  Sannino$^{\color{rossoCP3}{\spadesuit}}$}
\email{sannino@cp3-origins.net} 
\affiliation{\mbox{
$^{\color{rossoCP3}{\spadesuit}}$
$CP^3$-Origins \& the Danish IAS,
Univ. of Southern Denmark, Campusvej 55, DK-5230 Odense}}
\affiliation{
$^{\color{rossoCP3}{\vardiamondsuit}}$
Nordita,
KTH Royal Institute of Technology and Stockholm University,
Roslagstullsbacken 23, SE-10691 Stockholm, Sweden
}

\begin{abstract}
We elucidate and extend the conditions that map gauge-Yukawa theories at low energies into time-honoured gauged four-fermion interactions at high energies.
These compositeness conditions permit to investigate  theories of composite dynamics through gauge-Yukawa theories. 
Here we investigate whether perturbative gauge-Yukawa theories can have a strongly coupled limit at high-energy, that can be mapped into a four-fermion theory.
Interestingly, we are able to precisely carve out a region of the perturbative parameter space supporting such a composite limit. This has interesting implications on our current view on models of particle physics.
As a template model we use an $SU(N_C)$ gauge theory  with $N_F$ Dirac fermions transforming according to the fundamental representation of the gauge group. The fermions further interact with a gauge singlet complex $N_F\times N_F$ Higgs that ceases to be a physical degree of freedom at the ultraviolet composite scale, where it gives away to the four-fermion interactions.  We compute the hierarchy between the ultraviolet and infrared composite scales of the theory and show that they are naturally large and well separated. Our results show that some weakly coupled gauge-Yukawa theories can be viewed, in fact, as  composite theories. It is therefore tantalising to speculate that the standard model, with its phenomenological perturbative Higgs sector, could hide, in plain sight, a composite theory. 
\vskip.7cm
{\noindent \footnotesize Preprint: CP3-Origins-2015-20, DNRF90 \& DIAS-2015-20, Nordita-2015-68}
\end{abstract}

\maketitle

%\newpage
%\tableofcontents

\newpage
\section{The Composite Facet of Gauge Yukawa Theories}
\label{generalgy} 
Gauge-Yukawa theories are what make up our current best bet for a description of Nature at the smallest scales; the Standard Model of particle physics (SM). However, in order to resolve any of the standing problems in the SM, it must be extended. In this work we dedicate ourselves to the study of a general set of gauge-Yukawa theories, inspired by the global symmetries of QCD and the electroweak theory. Keeping the initial discussion general we will then argue that it is possible to reinterpret and make use of a certain class of gauge-Yukawa theories that are not ultraviolet safe. This amounts in having Landau poles in the couplings of the theory that cannot be tamed controllably at least within perturbation theory. Our results serve as starting point of nonperturbative first principle numerical studies beyond the conformal window of asymptotically free theories \cite{Sannino:2004qp,Dietrich:2006cm,Pica:2010xq}.  

A gauge-Yukawa theory can be described by a Lagrangian of the general form
\ea{
\label{simplegy}
\mathcal{L}&=-\frac{1}{4g^2} F_{\mu\nu}F^{\mu\nu}+ i\bar{\Psi}\; \Dslash\Psi +  D_\mu\Phi D^{\mu}\Phi^{\dag} +(Y\bar{\Psi}\Phi\Psi +h.c)- V(\Phi) \ , \\
V(\Phi)&={m_0}^2 \, \Phi^\dag\Phi +\lambda(\Phi^\dag\Phi)^2,
\label{pot}
}
where $Y$, $m_0^2$, and $\lambda$ may be matrices in the fermion  $\Psi$ and scalar $\Phi$ field space. To define the theory the appropriate gauge group(s) and corresponding representations must be specified, while the choice of the Yukawa and scalar couplings determine the global symmetries of the theory.
This class of theories are renormalizable and have been extensively studied. 
When quantum effects are considered and counter terms are added to remove the ultraviolet divergences, all of the terms in the Lagrangian Eq.~\eqref{simplegy} receive corrections. Focussing on the fermionic and scalar sector of the theory the changes to the Lagrangian are given by
\ea{
&  i\bar{\Psi}\; \Dslash\Psi \rightarrow i(1+\delta_{Z_\Psi})\bar{\Psi}\; \Dslash\Psi , 
  \quad D_\mu\Phi D^{\mu}\Phi^{\dag}\rightarrow (1+\delta_{Z_\Phi}) D_\mu\Phi D^{\mu}\Phi^{\dag} \ ,\quad 
  \label{renorm1}
  \\[2mm]
  &m_0^2 \rightarrow m_0^2 +\delta_{m^2} = m_r^2 ,\quad Y \rightarrow Y+\delta_Y = Y_r ,
  \quad \lambda \rightarrow \lambda+\delta_\lambda = \lambda_r, \quad 
    \label{renorm2}
}
where $\delta_{Z_\Phi}$ and $\delta_{Z_\Psi}
$  are the corrections from field-strength renormalization of the scalars and fermions. 
Through the renormalization procedure, a renormalization scale $\mu$ is introduced, and when the operators above change as the renormalization scale is varied, the theory moves along a renormalization group (RG) flow in the space of couplings. 
Defining the couplings at a given energy scale 
picks out a unique RG trajectory of the flow. Therefore in principle, a specific gauge-Yukawa theory has an infinite number of physically different paths in the RG flow. It can happen, however, that the theory possesses ultraviolet interacting fixed points that, depending on the dimension of the critical surface, can increase the predictivity of the theory \cite{Litim:2014uca,Litim:2015iea}\footnote{Complete asymptotically safe theories have novel thermodynamic properties \cite{Rischke:2015mea} and provide new ideas for model building \cite{Sannino:2014lxa,Nielsen:2015una}}.  This is, of course, also true for asymptotically free gauge-theories such as QCD where the infrared dynamics of the theory is known once the external group-theoretical parameters of the theory are given, such as the number of quark-flavors and colors.

To retain the canonical form of the renormalized Lagrangian, the field-strength renormalizations may be absorbed by a redefinition of the fields, $\Phi\rightarrow~\Phi/(1+~\delta_{Z_\Phi})^{1/2}$, and 
$ \Psi \rightarrow \Psi /(1+{\delta_{Z_\Psi}})^{1/2}$, giving:
\begin{equation}
\label{gyrenorm}
\mathcal{L}=-\frac{1}{4g^2}F_{\mu\nu}F^{\mu\nu} +  i\bar{\Psi}\; \Dslash\Psi + D_\mu\Phi D^{\mu}\Phi^{\dag} +(\tilde{Y}\bar{\Psi}\Phi\Psi +h.c)   - V(\Phi) \ ,
\end{equation} 
with 
\begin{equation}
V(\Phi)={m_\Phi^2}\Phi^\dag\Phi +\tilde{\lambda}(\Phi^\dag\Phi)^2 \ .
\end{equation} 
The renormalized canonical parameters are in terms of the renormalized non-canonical ones given by:
\begin{equation}
\tilde{Y}=\frac{Y_r}{(1+\delta_{ Z_\Psi})\sqrt{(1+\delta_{Z_\Phi})}}, \qquad m^2_\Phi= \frac{m_r^2}{(1+\delta_{Z_\Phi})}, \quad {\rm and} \quad \tilde{\lambda}=\frac{\lambda_r}{(1+\delta_{Z_\Phi})^2} \ .
\label{canonize}
\end{equation}
In standard perturbation theory the denominators in the above expressions can be taken to unity, such that to lowest order $\tilde{Y} = Y_r$, $m_\Phi^2 = m_r^2$ and $\tilde{\lambda}=\lambda_r$. However, this identification breaks down if at strong-coupling the field-strength renormalizations grow big.  This is the situation we would like to investigate.

In particular, we want to consider in this work gauge-Yukawa theories, where the scalars are composite fields appearing only below a certain energy scale $\Lambda_{\rm UV}$. Above that scale, one should recover a theory of only fermions and gauge bosons. This means that the scalars must cease to propagate at the scale $\Lambda_{\rm UV}$, and there we must set \mbox{$\delta_{Z_\Phi} =-1$}. 
This physical requirement on the scalar field translates into requirements for the scalar and Yukawa couplings as well as the mass term of the renormalized Lagrangian in Eq.~\eqref{gyrenorm}, which we call \emph{the compositeness conditions}. 
It is convenient to express these conditions in the following form:
\ea{
\lim_{\mu\rightarrow\Lambda_{\rm UV}} \tilde{Y}^{-2} = 0 \ ,
 \qquad \lim_{\mu\rightarrow\Lambda_{\rm UV}} \frac{\tilde{\lambda}}{\tilde{Y}^4} \approx 
 \frac{\lambda_r}{Y_r^4} \
 , \qquad
 \lim_{\mu\rightarrow\Lambda_{\rm UV}} \frac{m_\Phi^2}{\tilde{Y}^2}
 \approx \frac{m_r^2 }{Y_r^2} \ ,
 \label{conditions}
}
where by the limit an inverse transformation from Eq.~\eqref{gyrenorm} to Eq.~\eqref{simplegy} is implied at the scale $\Lambda_{\rm UV}$.
This transformation is necessary, because the canonical couplings diverge at the scale $\Lambda_{\rm UV}$. In perturbation theory such a divergence is associated with the occurrence of a Landau pole. 
The approximation used requires also that the fermion wave function renormalization correction does not spoil Eq.~\eqref{conditions}.

We show now a particularly important case, where these conditions are matched onto a purely fermionic gauge theory at the composite scale.  
Consider the case when $\lambda_r = 0$ at the scale $\Lambda_{\rm UV}$. 
The Lagrangian at the scale $\Lambda_{\rm UV}$ in this case reads:
\ea{
\mathcal{L}&=-\frac{1}{4g_r^2} F_{\mu\nu}F^{\mu\nu}+ i\bar{\Psi}\; \Dslash\Psi + (Y_r\bar{\Psi}\Phi\Psi +h.c)-{m_r^2}\Phi^\dag\Phi ,
}
where we assume that $\delta_{Z_\Psi} \ll 1$, or equivalently that the interactions are very weak at  $\Lambda_{\rm UV}$.
Since there is no kinetic term for the scalars, we may eliminate them via their equations of motion, and the resulting Lagrangian is
\ea{
\mathcal{L}&=-\frac{1}{4g_r^2} F_{\mu\nu}F^{\mu\nu}+ i\bar{\Psi}\; \Dslash\Psi + \frac{Y_r^2}{m_r^2}(\bar{\Psi}\Psi)^2 ,
}
which has the structure of a generalized gauged Nambu-Jona-Lasinio (gNJL) model~\cite{Nambu:1961fr}. The link between the four fermion theory described above and a low energy gauge-Yukawa theory was first demonstrated in \cite{Bardeen:1989ds}.
To connect the picture to the effective field theory language, we may choose as renormalization conditions $m_r^2(\Lambda_{\rm UV}) = \Lambda_{\rm UV}^2$ and $Y_r^2 (\Lambda_{\rm UV}) = G$, with $G$ being the  dimensionless four-fermion coupling. Then the above Lagrangian takes the form of the following effective field theory:
\ea{
\mathcal{L}&=-\frac{1}{4g_r^2} F_{\mu\nu}F^{\mu\nu}+ i\bar{\Psi}\; \Dslash\Psi + \frac{G}{\Lambda_{\rm UV}^2}(\bar{\Psi}\Psi)^2 \ .
}

The attentive reader would have realised that to derive the gNJL effective theory from the gauge-Yukawa system we used not only the compositeness conditions Eq.~\eqref{conditions} but also that $\lambda_r = 0$. It is therefore important to know when this requirement may be satisfied starting from the gauge-Yukawa theory.  
Consider the following limit: 
\ea{
\lim_{\mu\rightarrow\Lambda_{\rm UV}} \frac{\tilde{\lambda}}{\tilde{Y}^2} \approx 
\lim_{\mu\rightarrow\Lambda_{\rm UV}}  \frac{\lambda_r}{(1+ \delta_{Z_\Phi}) Y_r^2}
\ .
}
One observes that if $\lambda_r$ does not vanish at the composite scale the above quantity diverges at $\Lambda_{\rm UV}$. If, however,  $\lambda_r \to 0$ in this limit, the ratio of 
$\frac{\lambda_r}{(1+ \delta_{Z_\Phi}) }$ may go to a constant value, thus yielding
\ea{
\lim_{\mu\rightarrow\Lambda_{\rm UV}} \frac{\tilde{\lambda}}{\tilde{Y}^2}
= \rm constant \ .
\label{ncondition}}
This new condition will be added to the list of compositeness conditions given in Eq.~\eqref{conditions}, further reducing the number of gauge-Yukawa theories that may admit a composite realization of the gNJL-type.

The previous conditions are non-perturbative in nature and can be exploited to investigate also the correspondence between the two types of theories. In particular, as we shall see, the correspondence enables us to study  certain aspects of theories of composite dynamics through gauge-Yukawa theories that feature a RG region, where the theories can be treated perturbatively. 
This result  shows that weakly coupled gauge-Yukawa theories at some intermediate energy scale are, de facto, composite theories. It is therefore tantalising to speculate that the standard model with its perturbative Higgs sector could hide, in plain sight,  a composite theory. 

Beyond perturbation theory one can use first principle lattice studies \cite{Hasenfratz:1991it,Gerhold:2007yb,Gerhold:2011mx,Chu:2015nha,Catterall:2013koa} for which our results can be viewed exploratory in nature.  

We introduce a concrete example in Section \ref{template} that we use to elucidate the main points. It consists of an $SU(N_C)$ gauge theory featuring $N_F$ Dirac fermions transforming according to the fundamental representation of the gauge group. They further interact with a gauge-singlet $N_F \times N_F$ complex scalar field via Yukawa interactions that at intermediate energies self-interact. We show that it is possible to enforce the compositeness conditions in this theory while simultaneously discovering a controllable perturbative regime along the RG flow. This situation is similar to the SM, where at and around the electroweak scale all the couplings can be treated in perturbation theory. Because we have a clear perturbative regime, we divide the section in several subsections associated to different orders in perturbation theory. We show that the theory can admit a composite nature and furthermore estimate the ratio of the ultraviolet composite scale to the infrared chiral symmetry/confining scale as function of the parameters of the theory. We offer our conclusions in Section \ref{Conclusions}. A series of appendices contain detailed computations used to derive the results in the main text.

\section{The Composite Template}
\label{template}
We start with an $SU(N_C)$ gauge theory with $N_C>2$. The associated gauge fields $A^a_\mu$ have field strength $F^a_{\mu\nu}$ $(a=1,\cdots N_C^2 - 1)$. We add $N_F$ Dirac fermions $Q_i^c$ with $i=1,\cdots N_F$  and $c=1, \cdots N_C$ transforming according to the fundamental representation of $SU(N_C)$. The fermions further interact with an $N_F\times N_F$ complex scalar $H$. The fundamental interaction Lagrangian reads: 
\begin{equation}
\label{bosonized}
\mathcal{L} = - \frac{1}{2}\, \Tr \left[F^{\mu \nu} F_{\mu \nu}\right] + \Tr\left[
\overline{Q}\,  i\slashed{D}\, Q \right]  +\Tr\,\left[\partial_\mu H ^\dagger\, \partial^\mu H \right] + y\,\Tr\left[ \overline{Q}\, H\, Q  \right] - V\left[H \right]
\ ,\end{equation}
with $\Tr\left[ \overline{Q}\, H\, Q  \right]  = \Tr \left[\overline{Q}_L H Q_R + \overline{Q}_R H^{\dagger} Q_L\right]$ and 
\begin{equation}
\label{TreePotential}
V\left[H\right] =  m^2_H \, \Tr\left[H^\dagger H \right] + u\,\Tr\left[ H ^\dagger H H ^\dagger H  \right]  + v\,\left(\Tr\left[H ^\dagger H\right] \right)^2 \ . 
\end{equation}
%%% u1 --> v, u2--> u 

We trace over both color and flavour indices. This theory has been investigated in much detail recently in \cite{Antipin:2011ny,Antipin:2011aa,Antipin:2012kc,Antipin:2012sm,Antipin:2013pya} for a large number of interesting properties, not directly connected with compositeness.
It has been studied earlier in connection with top-quark condensate models in~\cite{Chivukula:1992pm,Bardeen:1993pj}, albeit in a different setup and limit that we here are taking.

The model has four classically marginal coupling constants given by the gauge coupling, the Yukawa coupling $y$, and the quartic scalar couplings; the single-trace coupling $u$ and the double-trace coupling $v$. From these we define new rescaled couplings, useful in the large $N_C$ and $N_F$ limit, which read
\begin{equation}\label{couplings}
\al g=\frac{g^2\,N_C}{(4\pi)^2}\,,\quad
\al y=\frac{y^{2}\,N_C}{(4\pi)^2}\,,\quad
\al u=\frac{{u}\,N_F}{(4\pi)^2}\,,\quad
\al v=\frac{{v}\,N^2_F}{(4\pi)^2}\, 
%, \quad M_H^2 = \frac{m_H^2 N_F}{(4\pi)^2} \ 
.
\end{equation}
These are the appropriately normalized couplings which enables us to study the Veneziano limit of the theory, where $N_F,N_C \rightarrow \infty$, while $N_F/N_C$ is kept constant. Note the additional power of $N_F$ in the definition of the scalar double-trace coupling, which makes $v/u$ go as $\al v / (\al u\,N_F)$.

The resulting compositeness conditions introduced in the previous section specialize to 
\begin{align}
 \lim_{\mu\rightarrow\Lambda_{\rm UV}} \al y^{-1} = 0 \ ,
 \qquad \lim_{\mu\rightarrow\Lambda_{\rm UV}} \frac{\al u}{\al y^2} =\lim_{\mu\rightarrow\Lambda_{\rm UV}}\frac{\al v}{\al y^2}  = 0 \ ,
 \qquad \lim_{\mu\rightarrow\Lambda_{\rm UV}}\frac{y^2}{m_H^2} = \frac{G}{\Lambda_{\rm UV}^2},
 \label{Compositeness}
 \end{align} 
where the last requirement gives the matching to the high energy four fermion theory.
The two first conditions can be investigated in any renormalization scheme, while the last one involving the mass, only applies to mass-dependent schemes. In mass-independent schemes there will be corrections to the right-hand-side of the latter condition~\cite{Bando:1991fv}, which are, however, unimportant to this work.
The matching to the high-energy theory is achieved in the following way:
At the scale $\Lambda_{\rm UV}$, where the couplings of the Lagrangian Eq.~\eqref{bosonized}
formally diverge, the theory should instead be rewritten through the transformations given in Eq.~\eqref{canonize}.
Assuming furthermore that
\ea{
 \lim_{\mu\rightarrow\Lambda_{\rm UV}} \frac{\al {u/v}}{\al y} = \rm constant \ .
}
as explained in the previous section, it then follows that the scalar sector of the theory is described by
\begin{align}
\mathcal{L}^H_{\text{Composite}} &= \sqrt{G} \, \Tr \left[\overline{Q}_L H Q_R + \overline{Q}_R H^{\dagger} Q_L\right] - \Lambda_{\rm UV}^2 \, \Tr\left[H^\dagger H \right] 
\ ,
\end{align}
where the fields $Q_{L/R}$ and $H$ now are the inversely transformed ones of Eq.~\eqref{renorm1}. The renormalized mass parameter and Yukawa coupling are the inversely transformed ones defined in Eq.~ \eqref{renorm2}, where
the renormalization conditions identifying them with the cutoff and the four-fermion coupling was imposed. 
By eliminating the auxiliary scalar degrees of freedom through their equation of motion, one 
obtains the four-fermion interaction\footnote{By using a Fierz identity, this can be recast into the form $G/\Lambda_{\rm UV}^2(\overline{Q}_{L }^{i c}Q^j_{R c})(\overline{Q}_{R j}^{c'}Q_{L i c'})$
  .}:
\ea{
\mathcal{L}^H_{\text{Composite}} 
= \frac{2G}{\Lambda_{\rm UV}^2} \, \Tr\left[\overline{Q}_L T^a Q_R \right] \Tr \left[\overline{Q}_R T^a Q_L\right] \ ,
}
Here $T^a$ was introduced through $H= h^a T^a$, with $a=0,1,\hdots, N_F^2-1$ which are the the generators of $SU(N_F)$, while $T^0=\tfrac{1}{\sqrt{2N_F}} \mathbb{1}$. The normalization used is $\Tr T^a T^b = \frac{1}{2} \delta^{ab}$.

We are now ready to provide a consistent renormalization group investigation of this gauge-Yukawa system  superimposed with the compositeness conditions derived above.
The renormalization group flow of a gauge-Yukawa theory arranges itself in a particular pattern in perturbation theory. As shown in \cite{Antipin:2013sga} the beta functions of these theories in mass-independent schemes abide the Weyl consistency conditions \cite{Osborn:1991gm}. These conditions have been further tested in \cite{Jack:2014pua}. These dictate a specific counting scheme to correctly take into account higher-order corrections. 
This counting scheme can in perturbation theory also be understood through the general pattern for the perturbative beta functions of the dimensionless couplings in mass-independent schemes:
\ea{
\beta_g &= \beta_g^{(1)}(g) + \beta_g^{(2)}(g,y) + \beta_g^{(3)}(g,y, \lambda) + \cdots \ ,\\
\beta_y &= \beta_y^{(1)}(g,y) + \beta_y^{(2)}(g,y, \lambda) + \cdots \ ,
\\
\beta_\lambda &= \beta_\lambda^{(1)}(\lambda, g ,y) + \cdots \ ,
}
where the superscripts denote the loop order of the terms and the parenthesis shows which couplings they depend on.
This pattern is completely general, and shows that one may consider the running of the gauge coupling at 1-loop consistently without taking into account the running of Yukawa and the quartic couplings (leading order). Likewise one may analyze the two-loop running of the gauge coupling taking into account the one-loop running of the Yukawa consistently without taking into account the running of the quartics (next-to-leading order). 
At three loops, running of all couplings must be taken into account and the lowest consistent counting order is 3-2-1 loops in the gauge-Yukawa-quartic beta functions (next-to-next-to-leading order).
The Weyl consistency conditions, in fact, dictate that this is the only consistent counting scheme.
We will in this sense analyze the leading, next-to-leading and next-to-next-leading order corrections to the RG flow and their physical implications on the four-fermion theory described above. In particular, we will compute the distance in energy between the composite scale and the confinement scale of the theory, and show that large hierarchies are not only possible to establish, but seems to be a clear feature of these theories.

\subsection{Leading order and weak compositeness conditions}
The leading order analysis is an over simplified case, which is not able to capture the composite nature of gauge-Yukawa theories. Nevertheless, we make a leading order analysis in this section for completeness, since it allows us to define the infrared scale and furthermore provides a pedagogic step towards the following sections.

To the leading order one needs only to consider the gauge beta function at one-loop which reads:
\begin{equation}
\beta_g =  \partial_t \alpha_g =  - \beta_0 \al g^2 = - \frac{2}{3}\alpha_g^2\left({11}-2\frac{N_F}{N_C} \right) \ .
\end{equation}
The Veneziano limit allows us to further take $N_F/N_C$ to be any real nonnegative number, called $x = N_F/N_C$. Depending on the number of flavours,  the double zero at $\al g = 0$ can either be an infrared or an ultraviolet gaussian fixed point. The second case is also known as asymptotic freedom. In the first case the ultraviolet theory is not well defined unless higher orders introduce an interacting ultraviolet fixed point, in which case the theory becomes asymptotically safe \cite{Litim:2014uca,Litim:2015iea}. 

\begin{figure}[bt]
\begin{center}
\includegraphics[width=0.5\textwidth]{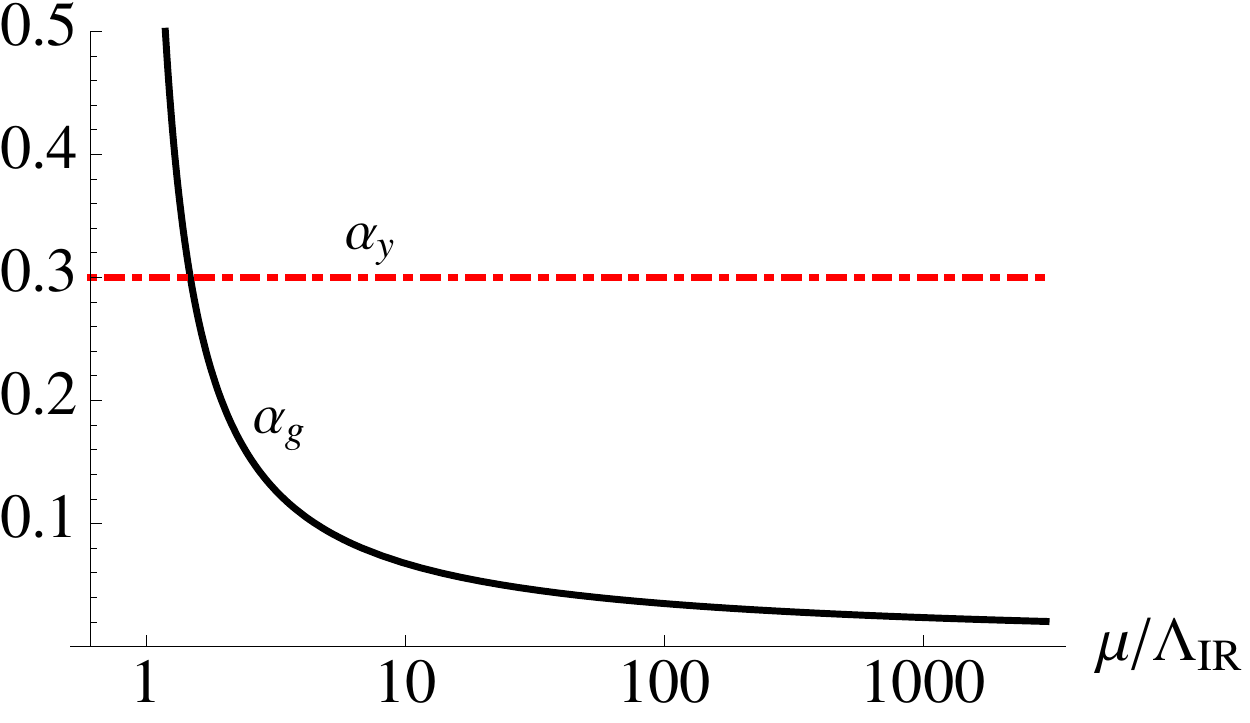}
\caption{Renormalization group evolution for the lowest order analysis, where the Yukawa coupling is constant and nonzero, while the gauge coupling runs. We have here defined the scale $\Lambda_{\rm IR}$ such that $\alpha_g(\Lambda_{\rm IR})=1$. The scalar quartic couplings, not included here, are also constants to this order. }
\label{warmupgraph}
\end{center}
\end{figure}

Here we consider the case in which the theory is asymptotically free. 
This restriction allows us to assume that the wave-function renormalization of the fermions will stay small near the composite scale, since they are at one-loop produced by gauge interactions.
As explained in the previous section, for consistency we should not consider the running of the  scalar and Yukawa couplings at this order. 
According to the compositeness conditions Eq.~\eqref{Compositeness} we should have
\begin{equation}
\lim_{\mu\rightarrow\Lambda_{\rm UV}}{\al y}^{-2} = 0  \ .
\label{weak}
\end{equation}
To this order, a constant and formally divergent $\al y$ is thus required.
This is in clear tension with perturbation theory.
Given that we want to avoid an uncontrollable nonperturbative analysis, to the leading order we therefore must take another approach by enforcing instead a weaker version of the compositeness conditions: Assuming that we are describing a four-fermion theory at a mass scale, which is at least a few times below the composite scale, we may consider $\alpha_y$ simply to some constant value smaller than one, as depicted in Fig.~\ref{warmupgraph}, to ensure validity of the pertubative analysis.
We shall see that when next to leading order corrections are taken into account, this assumption is valid, since the Yukawa coupling will naturally grow at high energy and what we are describing here are boundary conditions in an energy range, where the Yukawa coupling is small enough for perturbation theory to hold. For the scalar self-interactions we assume a similar behaviour.

From this first oversimplified analysis one concludes that the asymptotically free theory develops a mass gap associated with the divergence of the gauge coupling at low energies.  
At these energies chiral symmetry breaks leading to the formation of the nonperturbative condensate 
\begin{equation}
\langle \overline{Q} Q\rangle \propto \Lambda_{\rm IR}^3 \ .
\end{equation}
The scale $\Lambda_{\rm IR}$ can be estimated to be (cf. Eq.~\eqref{strongscale} in the appendix):
\ea{
\Lambda_{\rm IR} = \mu_0 \exp \left (-\frac{1}{\beta_0 \al g(0)} \right ) \ .
}
{This estimate is insensitive to the perturbative corrections from the Yukawa and scalar sectors, which contribute only at higher orders. 
Thus if the Yukawa and scalar sectors stay perturbative in the IR, the above expression provides a good estimate of the IR strong scale of the fully dynamical gauge-Yukawa theory.}

\subsection{Next-to-leading order analysis: The rise of the Yukawa coupling}

For the next order in perturbation theory  one needs to go to two loops in gauge coupling and one loop in the Yukawa, while the running of the scalar couplings are still not relevant. To this order, therefore, the Yukawa coupling is no longer a constant and its running and consequent back-reaction on the gauge coupling are important. We have
\footnote{{The beta functions are in the $\overline{\text{MS}}$-scheme  
\cite{Luo:2002ti,Machacek:1983tz,Machacek:1983fi,Machacek:1984zw,Pickering:2001aq,Mihaila:2012pz}. It would also be interesting to investigate the compositeness conditions in other renormalisation schemes such as the momentum subtraction scheme \cite{Gracey:2015uaa,Ryttov:2014nda,Ryttov:2013ura,Ryttov:2012ur}, since different schemes can be more or less suitable to explore different facets of gauge-Yukawa theories.}}: 
\begin{eqnarray}
\beta_g & = &  -\frac{2}{3}\alpha_g^2 \left[  \left({11}- 2\frac{N_F}{N_C}\right)  + \left({34} -\frac{N_F}{N_C}\left\{10+3\frac{N_C^2-1}{N_C^2}\right\}\right)\alpha_g + 3\frac{N_F^2}{N_C^2}\alpha_y \right]\ ,\\  
\beta_y & = & \alpha_y\left[2\left(1+\frac{N_F}{N_C}\right)\alpha_y -6\frac{N_C^2-1}{N_C^2}\alpha_g \right]\ .
\end{eqnarray}
Working in the Veneziano limit by defining $x = N_F/N_C$ at large $N_F$ and $N_C$ yields: 
\begin{eqnarray}
\label{venegauge2}
\beta_g & = &  -\frac{2}{3}\alpha_g^2 \left[  \left({11}- 2x\right)  + \left({34} -13x\right)\alpha_g + 3x^2\alpha_y \right]\ ,\\  
\beta_y & = & 2\alpha_y\left[\left(1+x\right)\alpha_y -3\alpha_g \right]\ .
\label{veneyuk1}
\end{eqnarray}
We restrict $x < 11/2$, ensuring asymptotic freedom for the gauge coupling. 
In the absence of the Yukawa interactions, a well known interacting infrared fixed point emerges at 
\begin{equation}
\alpha_g^{\ast} = \frac{11-2x}{13 x - 34} \ , \qquad {\rm for}  \qquad  \frac{34}{13}< x < \frac{11}{2}\ , \quad {\rm and} \quad \alpha_y = 0\ . 
\end{equation}
For $x$ very close to $11/2$ this is the Banks-Zaks perturbative infrared fixed point. This fixed point, however, disappears  in the presence of the Yukawa interactions \footnote{
If the infrared fixed point should exist, it should be a fixed point also for the Yukawa interactions. By setting $\beta_y = 0$  we derive $\alpha_y= \tfrac{3}{1+x}\alpha_g$ which can be substituted in $\beta_g$ yielding: 
\begin{equation}
\beta_g  \rightarrow   -\frac{2}{3}\alpha_g^2 \left[  \left({11}- 2x\right)  + \left({34} -13x + 9 \frac{x^2}{1+x} \right) \alpha_g\right]\ , \nonumber
\end{equation}
showing that the presence of the Yukawa has eliminated the possibility of the infrared fixed point. }. Therefore the next-to-leading-order effects on the gauge beta function strengthens the infrared QCD-like behaviour of the theory. 

\begin{figure}[bt]
\begin{center}
\includegraphics[width=0.45\textwidth]{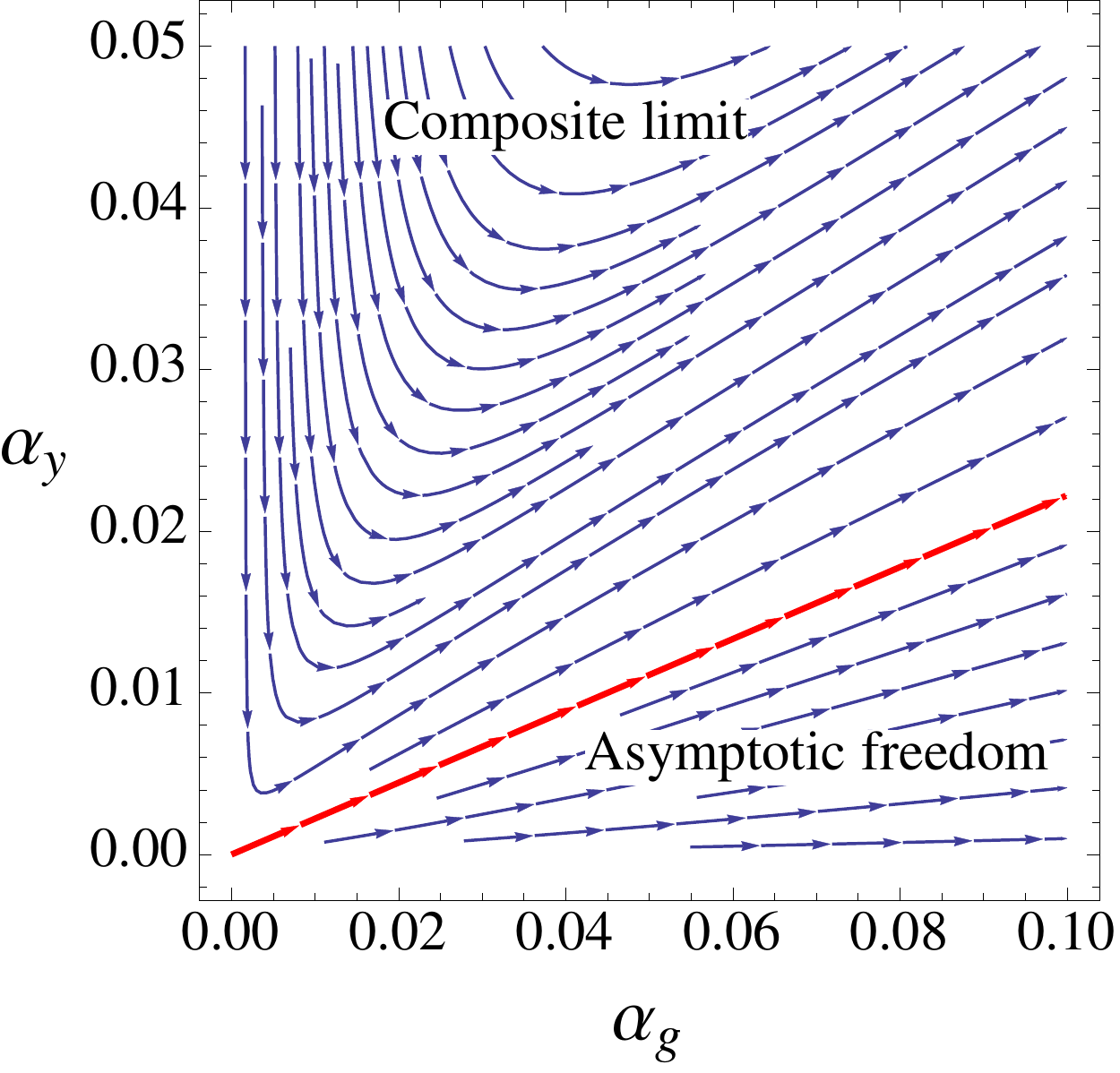}
\vspace{-3mm}
\caption{The RG flow in the ($\alpha_g,\alpha_y$) plane for $x = 2$. Two distinct phases are present. The red trajectory indicates the phase boundary estimated from the one loop beta functions: $\frac{\alpha_y}{\alpha_g}=\frac{2(x-1)}{3(x+1)}$.}
\label{gyflow}
\vspace{-3mm}
\end{center}
\end{figure}

 The RG flow of the gauge-Yukawa system for $x=2$ is shown in Fig.~\ref{gyflow}. 
The arrows in the figure shows the flow from the ultraviolet (UV) to the infrared (IR) regime.
In the UV two distinct phases form. The boundary between these two phases is approximately given by 
\ea{
\frac{\alpha_y}{\alpha_g}=\frac{2(x-1)}{3(x+1)} \ ,
\label{ayagboundary}
} 
which is determined by the one-loop beta functions in both couplings (cf. Eq.~\eqref{compcond} in the appendix). Below the red trajectory both couplings are asymptotically free meaning that the theory is non-interacting and well defined in the UV. This RG region, therefore, does not support a composite limit of theory. The composite limit emerges in the RG region above the red trajectory, where the Yukawa coupling diverges in the UV, thus allowing the compositeness conditions given in Eq.~\eqref{Compositeness} to be satisfied. 
We notice that the boundary Eq.~\eqref{ayagboundary} for $x\leq 1$ is outside the physical parameter space of the couplings. Therefore the composite limit is supported by the entire perturbative region of the physical space of couplings, i.e. the Yukawa coupling will also diverge in the UV. Thus $x=1$ defines a boundary in the external parameter space.

\begin{figure}[b]
\begin{center}
\includegraphics[width=0.5\textwidth]{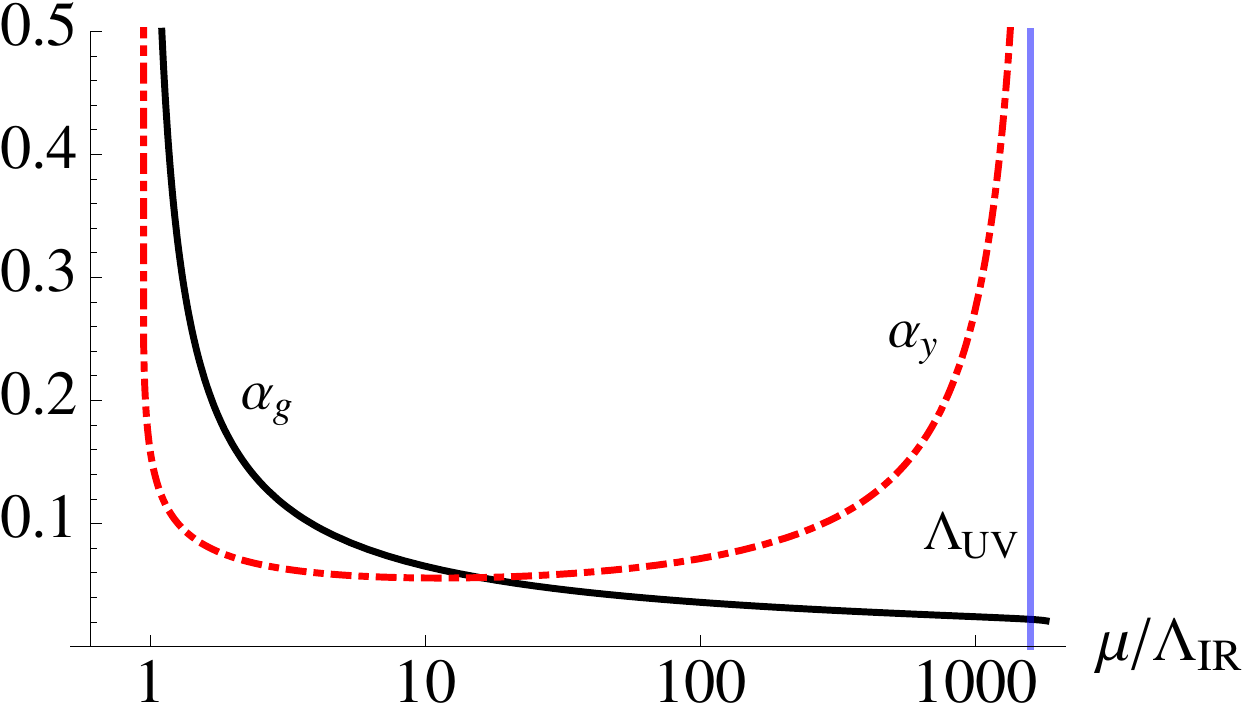}
\caption{Renormalization group evolution for the next to leading order analysis, where the Yukawa and gauge couplings run. We define the scale $\Lambda_{\rm IR}$ and  $\Lambda_{\rm UV}$ such that $\alpha_g (\Lambda_{\rm IR})=\alpha_y(\Lambda_{\rm UV})=1$. The scalar quartic couplings are constants to this order and are not included here.}
\label{2loopgraph}
\end{center}
\end{figure}

We show in Fig.~\ref{2loopgraph}, again for $x=2$, the actual running of the two couplings in the composite region for one particular RG trajectory.
Considering the flow from UV to IR, initially $\al y \gg \al g$ due to the compositeness condition. Since $\al y$ will decrease, while $\al g$ increases towards the IR, at some intermediate scale $\mu_0$, their values cross, and once $3 \al g > (1+x)\al y$, the sign of the Yukawa beta function changes, making it grow again in the deep IR.
This growth of $\alpha_y$ in the IR is therefore at most as fast as $3\alpha_g/(1+x)$.

The composite scale $\Lambda_{\rm UV}$ is identified with the Landau pole in the  Yukawa coupling. We will fix our perturbative initial conditions at the crossing scale $\mu_0$, and we ensure perturbation theory to be valid by requiring $\alpha_g(\mu_0) =\alpha_y (\mu_0) =C \ll 1$.  This condition is for any $x$ consistently above the boundary Eq.~\eqref{ayagboundary}, ensuring the theory to be in the composite phase.

It is interesting to study  the hierarchy between the  composite scale   and the chiral symmetry breaking one, as a function of both $C$ and $x$. At the one loop level , in both the gauge and Yukawa coupling, we can estimate it analytically to be
\begin{equation}
 \log \left(\frac{\Lambda_{\rm UV}}{\Lambda_{\rm IR}}\right)
= \frac{3\left(1+\frac{\alpha_g(\mu_0)}{\alpha_y(\mu_0)}\frac{2(1-x)}{3(1+x)}\right)^{\frac{11-2x}{2(1-x)}}}{2(11-2x) \alpha_g(\mu_0)} \, .
\label{scaleratio}
\end{equation}
The expression is well-defined for any $x$, and it takes the following simple form at $x=1$:
\ea{
\lim_{x\to1} \log \left(\frac{\Lambda_{\rm UV}}{\Lambda_{\rm IR}}\right)
= \frac{1}{3 \al g(\mu_0)} \lim_{x\to 1}
\left(1+\frac{\alpha_g(\mu_0)}{\alpha_y(\mu_0)}\frac{2(1-x)}{3(1+x)}\right)^{\frac{11-2x}{2(1-x)}}
= \frac{\exp \left (\frac{3\al g (\mu_0)}{2\al y (\mu_0)}\right )}{3 \al g(\mu_0)}   \ .
}
To set the initial values of the couplings we  will use  $\alpha_g(\mu_0) =\alpha_y (\mu_0) =C $ since in the composite phase there will always be a $\mu_0$ such that this condition is fulfilled. 

\begin{figure}[b]
\begin{center}
\subfigure[ \, $x$ is varied while $C=0.1$]{
\includegraphics[width=0.45\textwidth]{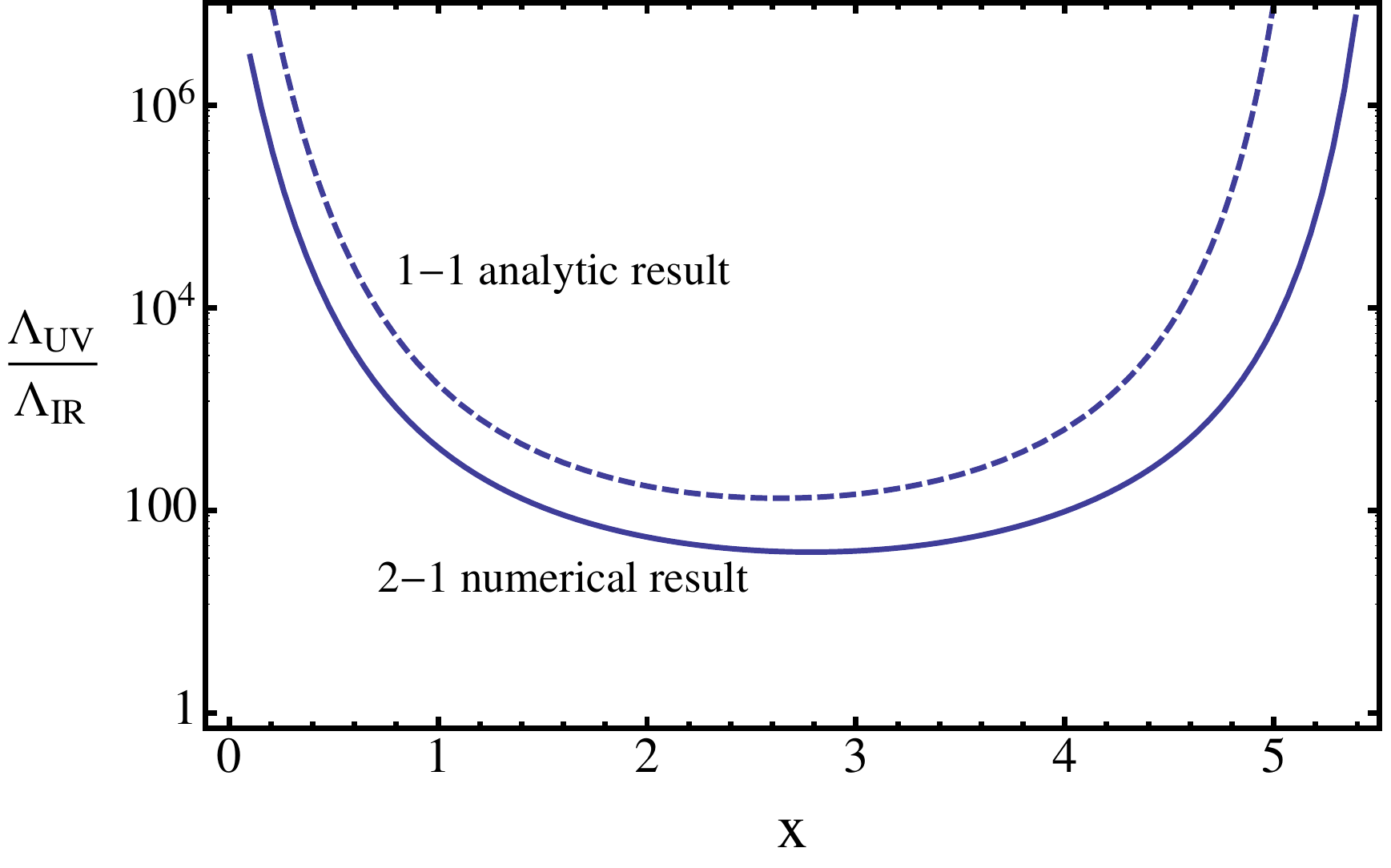}
\label{varyx}}
\quad 
\subfigure[ \,$C$ is varied while $x=2.5$]{
\includegraphics[width=0.45\textwidth]{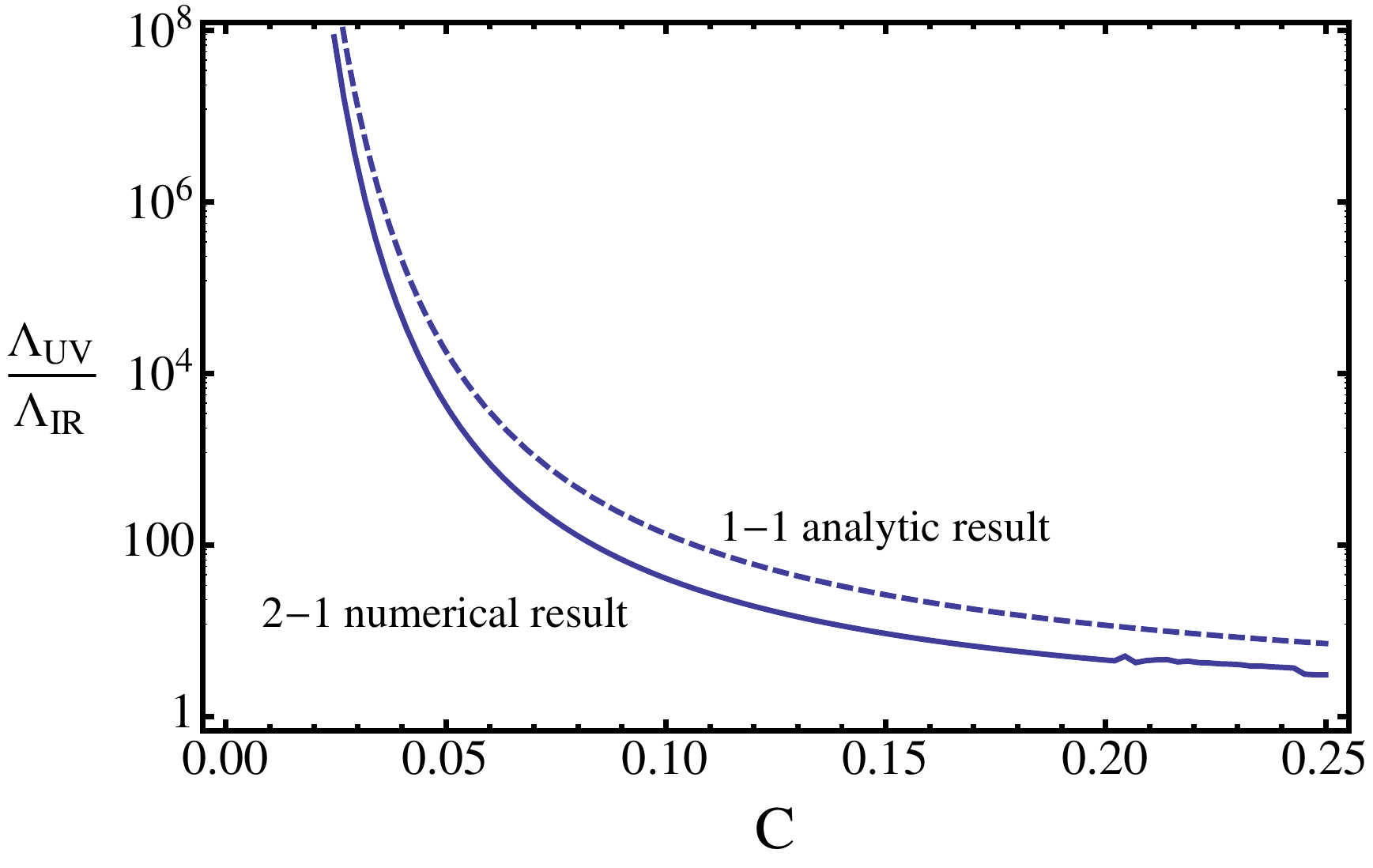}
\label{varyC}}
\caption{The ratio between the scale of UV compositeness $(\Lambda_{\rm UV})$ and the one associated to infrared gauge coupling divergence $(\Lambda_{\rm IR})$, as a function of the number of flavors/colors, parametrized by $x$ and the common value, $C$, of the gauge and Yukawa coupling at the scale where they are equal. We vary $x$ in \subref{varyx} and $C$ in \subref{varyC}. The one loop estimate is presented in dashes while the NLO perturbative (2-1) numerical result is represented by the solid curve}
\label{ratio}
\end{center}
\end{figure}

In Fig.~\ref{ratio} we compare the approximate analytical one-loop result with the next-to-leading order numerical calculation.  To numerically estimate the value of the IR(UV) scale we use the approximate relation $\alpha_{g(y)}(\Lambda_{IR(UV)})=1$. 

The ratio increases for small and large values of $x$ for a fixed value of $C$. This is because for small $x$ the first coefficient of the Yukawa beta function decreases, de facto, slowing the runaway behavior of the associated coupling in the UV. For large $x$, instead, the ratio becomes large since we are nearing the limit where asymptotic freedom is lost for the gauge coupling. Consequently the infrared scale is approaching zero. 

Additionally, we explore the influence of the chosen value of the couplings $\alpha_y=\alpha_g=C$ at the scale where they are equal. Setting $x=2.5$ associated to a region of $x$ that does not influence dramatically the ratio $\Lambda_{\rm UV}/ \Lambda_{\rm IR}$, as it is clear from Fig.~\ref{varyx}, we vary the value $C$ and plot again the ratio in  Fig.~\ref{varyC}.
As one might have expected, smaller values of the couplings lead to a larger ratio of the scales since more RG running is needed to reach the UV and IR scales where the couplings become nonperturbative. 
We also observe that the approximate one-loop result overestimates the ratio. 
 
 Another interesting feature is that for the theory to remain perturbative in an intermediate regime, say $C< 0.1$, the ratio of the scales, as function of $x$, cannot be too small, and typically should be larger than $100$, implying a hierarchy of scales of at least two orders of magnitudes. This has interesting phenomenological consequences which will be discussed later.  
 
 It is straightforward to see that the compositeness conditions in Eq.~\eqref{Compositeness} are satisfied to this order. The conditions for the scalar sector are satisfied by imposing the weaker version of the conditions discussed in the previous leading-order case. Following that reasoning, thus to the next-to-leading order in perturbation theory, we have shown that gauge-Yukawa theories can be naturally viewed as stemming from a compositeness paradigm for a wide region of the RG phase diagram, e.g. the one in Fig.~\ref{gyflow}.

\subsection{Next-to-Next-to-leading order: The awakening of the scalars}
In the previous sections, we were able to draw a consistent picture of compositeness in the gauge-Yukawa sector, and we were furthermore able to provide estimates for the hierarchy between the ultraviolet composite scale and the infrared confinement scale.
From the ultraviolet theory point-of-view, the scalars are merely auxiliary fields. For consistency of the analysis in the previous sections, they should therefore not play any physical role. In this section, we investigate the influence of the scalars on the above results, and provide the needed constraints on the scalar coupling phase space, needed to ensure consistency of the previous analysis.

The next order in the RG analysis requires the one loop beta functions for the quartic couplings, the two loop terms in the Yukawa beta function, and the three loop terms in the gauge beta function. This system of RG equations obeys the Weyl consistency conditions and reflects the back reaction from the scalars on the running of the Yukawa coupling, which in turn back reacts on the gauge coupling. Since the scalars do not carry gauge charge, they do not contribute to the three-loop terms for the gauge coupling.
Additionally, since we are considering a mass-independent renormalization scheme, we can independently take into account the running of the mass, where one-loop is also sufficient.
In the Veneziano limit, the beta functions to this order read~\cite{Antipin:2012kc,Antipin:2013pya}:
\ea{
\label{venegauge3}
\beta_g  = &  -\frac{2}{3}\alpha_g^2 \left[  \left({11}- 2x\right)  + \left({34} -13x\right)\alpha_g + 3x^2\alpha_y +\frac{81x^2}{4}\alpha_g\alpha_y  
\right. \nonumber \\
 & \qquad \quad \  \left. 
  -\frac{3x^2(7+6x)}{4}\alpha_y^2 +\frac{2857+112x^2-1709x}{18} \alpha_g^2\right]\ ,\\  
\beta_y  = & 2\alpha_y\left[\left(1+x\right)\alpha_y -3\alpha_g + (8x+5)\alpha_g\alpha_y +\frac{20x-203}{6}\alpha_g^2 
-8x\alpha_u-\frac{x(x+12)}{2}\alpha_y^2+4\alpha_u^2\right]\ ,
\label{veneyuk2}
}
and for the scalar sector
\begin{eqnarray}
\beta_u & = & 4\left[2\alpha_u^2 +\alpha_u\alpha_y-\frac{x}{2}\alpha_y^2 \right] \ ,
\label{veneau1}
\\
\beta_v & = & 4\left[\alpha_v^2 +4\alpha_u\alpha_v+3\alpha_u^2+\alpha_v\alpha_y \right]\ ,
\\
 \beta_{m_H^2} &= &\partial_t m_H^2 = 4 m_H^2 [  \alpha_y +  \alpha_v + 2 \al u ] \ .
\label{veneav1}
\end{eqnarray}

The RG structure of this theory is quite rich and has been intensively studied in recent years~\cite{Antipin:2011ny,Antipin:2011aa,Antipin:2012kc,Antipin:2012sm,Antipin:2013pya,Litim:2014uca,Litim:2015iea,Antipin:2014mga}.
Here, we are interested in a new point-of-view, which concerns compositeness.
 
In this section we must show that the scalar self-interactions can be consistent with the compositeness picture emerged above and driven, so far, by the Yukawa interactions. Specifically, considering as in the above analysis an intermediate RG scale $\mu_0$, where perturbation theory is well-defined, we have to ensure that the scalar couplings stay perturbative up to the composite scale, where they furthermore have to satisfy the compositeness conditions given in Eq.~\eqref{conditions} and Eq.~\eqref{ncondition}. The reason for this requirement is that if the scalar couplings would grow strong before the composite scale, the analysis of the previous sections would be invalidated. 

There are two other issues which may arise; the first is that according to the mass-independent scheme, the scalars remain dynamical as long as $m_H(\mu) < \mu$.
For $m_H(\mu) = \mu$, the scalars will decouple before reaching the scale where they should be seen as auxiliary fields, and therefore this situation should be avoided. 
The second issue arises when the effective potential develops a global minimum away from the origin due to quantum corrections. 
For consistency of our analysis, this has to be avoided between the scales $\mu_0$ and $\Lambda_{\rm UV}$,
since the vacuum expectation value of the scalar fields was earlier assumed to be zero in the analysis of the compositeness condition on the Yukawa coupling and in the calculation of the scale hierarchy.
However, at lower scales there is no inconsistency of having a symmetry breaking through the scalar sector, rather than the gauge sector.
This would correspond to another interesting possibility that we are, however, not considering here.
 
To summarize, the aim of this section is to understand and provide the criteria under which:
\begin{enumerate}
\item
The scalar sector stays perturbative up to the composite scale, where it furthermore must satisfy the compositeness conditions.

\item The scalars do not decouple before the infrared confinement scale.

\item
The minimum of the effective potential at the origin remains stable under quantum corrections between the composite and confinement scales.
\end{enumerate}

We will now demonstrate that there is a subset of theories which do obey the above three constraints on the scalar sector.
First of all, we need to ensure that there is no Landau pole in the scalar couplings between $\mu_0$ and $\Lambda_{\rm UV}$. 
To lowest order in perturbation theory, we have shown in the appendix (cf. Eq.~\eqref{tu}, \eqref{aubounds} and \eqref{tv3}) that
the initial conditions on the scalar couplings must satisfy the following inequality, to not become strong at intermediate scales:
\ea{
\Big \{ |\al u(\mu_0) | \ , \ 2 |\al v (\mu_0) | \Big \}  < \frac{2(11-2x)  }{24} \frac{C %\al g(0)
 }{\left (1- \frac{2}{3}\frac{x-1}{x+1} 
 %\frac{\al g(0)}{\al y(0)} 
 \right )^{\frac{11-2x}{2(1-x)}} - 1 } \ ,
 \label{auavcon1}
}
where we used the renormalization condition of the previous section $\al g (\mu_0) = \al y(\mu_0) = C$.

There is another subtle effect, which can lead to a Landau pole, due to tangential divergence, as explained in the appendix (cf. Eq.~\eqref{auintermediate}-\eqref{aubound2}). Here we can in the general case at best impose an overconstraint inequality, ensuring no Landau poles. For the $\al u$ coupling it reads:
\ea{
-1-\sqrt{1+4x}
&<  4 \frac{\al u(\mu_0)}{ C} < 
-1+\sqrt{1+4x} \ .
\label{aucon2}
}
For the $\al v$ coupling, the situation is more complex (cf. Eq.~\eqref{avintermediate}). The following constraints, however, will ensure no Landau poles at intermediate scales:
\ea{
\al v(\mu_0) &> \frac{ \al y(\mu_0) + 4 \al u(\mu_0)}{2} \left ( -1 + \sqrt{1- \frac{12}{\left(4 + \frac{\al y(\mu_0)}{\al u(\mu_0)}\right )^2}} \right ) \ ,
\label{avvsau1}
\\
\al v(\mu_0) &< \frac{ \al y(\mu_0) + 4 \al u(\mu_0)}{2} \left ( -1 - \sqrt{1- \frac{12}{\left(4 + \frac{\al y(\mu_0)}{\al u(\mu_0)}\right )^2}} \right ) \ .
\label{avvsau2}
}
If $\al u(\mu_0)$ is negative, the additional constraint, $\left(4 + \frac{\al y(\mu_0)}{\al u(\mu_0)}\right )^2 > 12$, must be imposed, which can be expressed more clearly as:
\ea{ 
\al u(\mu_0) >-\frac{\al y(\mu_0)}{ 4 + \sqrt{12}} \approx -0.13 \, \al y(\mu_0)
 \quad \text{and} \quad 
\al u(\mu_0) < \frac{\al y(\mu_0)}{ \sqrt{12}-4} \approx -1.87 \, \al y(\mu_0) \ .
\label{aucon3}
}
To ensure that there are no Landau poles in the infrared regime, before the confinement scales, similar constraints can be put, which are also provided in the appendix (cf. Eq.~\eqref{aubounds} and \eqref{tv3}).

At high scales the coupling $\al v$ may also exhibit tangential divergence, as explained in the appendix (cf. Eq.~\eqref{Dconstraint}). This is avoided by imposing the following constraint on the theory parameters:
\ea{
x > - 4 + 3 \sqrt{3} - \sqrt{6\left (7-4\sqrt{3} \right )} \approx 0.54 \ . 
\label{last}
}
The constraint does not depend on the initial perturbative values of the scalar couplings and must be satisfied regardless. Thus we can conclude that for $x < 0.54$, the perturbative theory cannot show a composite nature of the type we are considering, but for any other values $0.54 < x < 5.5$, there are well defined regions where compositeness is expected.

When these constraints are satisfied, the only Landau pole appearing in the UV regime is the one driven by the Yukawa coupling. It is then clear that in perturbation theory the running of the scalar couplings at the composite scale may only diverge as fast as the Yukawa coupling, and thus the extra condition in Eq.~\eqref{ncondition}, in agreement with an NJL-type four-fermion theory interpretation, is automatically satisfied. In the appendix we have furthermore showed that the value of Eq.~\eqref{ncondition} are at one-loop exactly fixed by the theory parameters, and independent of the initial values of the couplings (cf. Eq.~\eqref{RyutL} and \eqref{RyvtL}). In particular, near the composite scale the sign of $\al u$ is always negative, while the sign of $\al v$ is always positive. The consequence of this on the stability of the potential will be analyzed at the end.

\begin{figure}[bt]
\begin{center}
\includegraphics[width=0.5\textwidth]{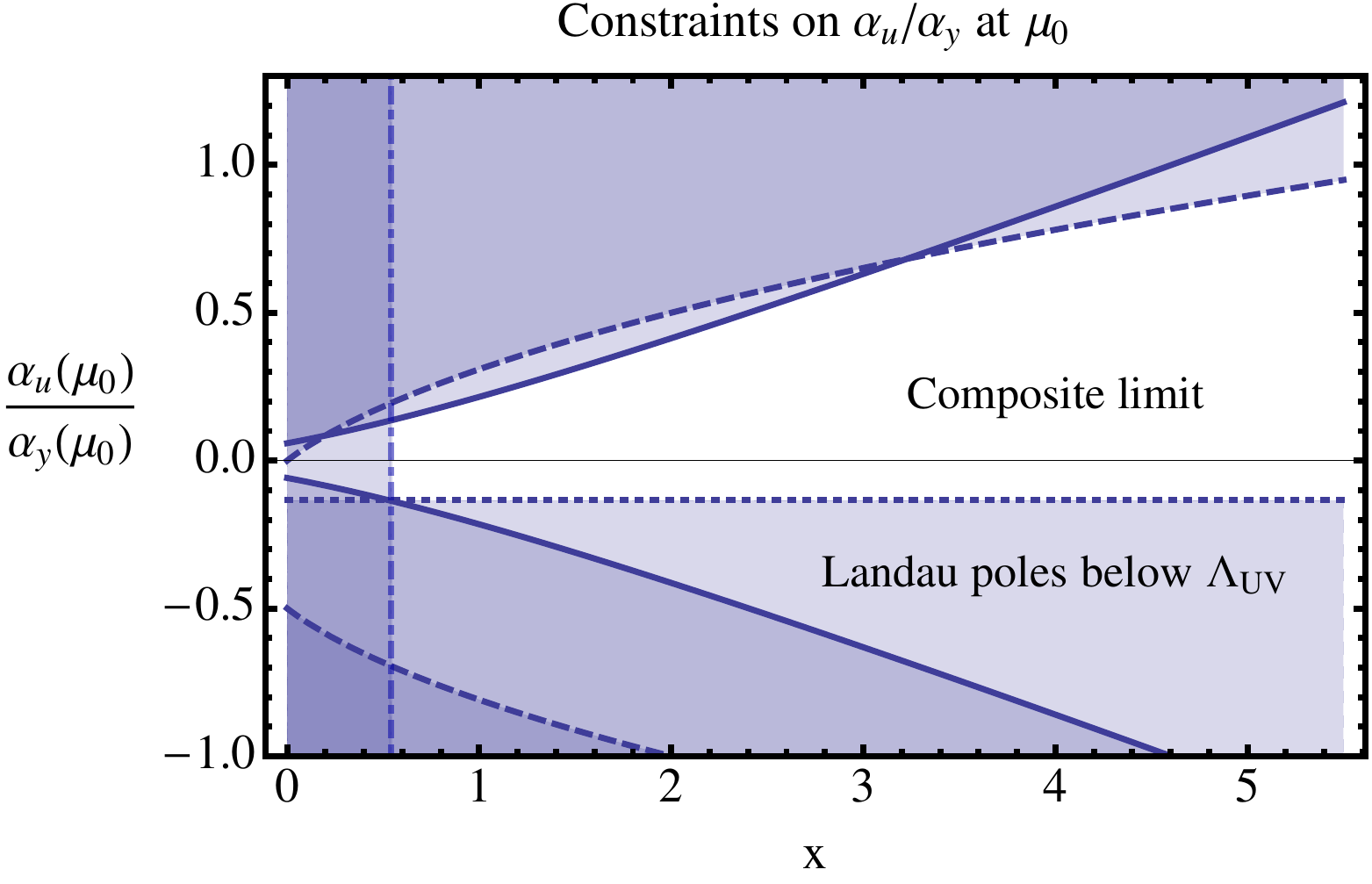}
\caption{The constraints on $\alpha_u(\mu_0)$ in terms of $\alpha_y(\mu_0)$ for different values of $x$. The strong coupling constraint Eq.~\eqref{auavcon1} is shown in solid, while the perturbative tangential divergence constraints on $\al u$ Eq.~\eqref{aucon2}, 
and on $\al v$ Eq.~\eqref{aucon3}-\eqref{last} are shown in dashes, dots, and dotdashes, respectively. }
\label{constau}
\end{center}
\end{figure}

Intuitive understanding of the constraints for the quartic couplings may most easily be obtained from a visualization, and in Fig.~\ref{constau} we display the constraints Eq.~\eqref{auavcon1}, \eqref{aucon2}, \eqref{aucon3} and \eqref{last} in terms of the ratio $\frac{\alpha_u(\mu_0)}{\alpha_y(\mu_0)}$.
From the figure we see that, although the absence of unwanted Landau poles is strongly constraining the parameter space, a range of initial values for $\alpha_u$ is still consistent with the composite picture. We note in particular that the quartic coupling $\alpha_u$ is always constrained to be smaller than the Yukawa coupling, and that for any $x$, the coupling $\al u$ cannot be smaller than $- 0.13 \al y$.

\begin{figure}[bt]
\begin{center}
\subfigure[\, 
$x$-independent bounds on compositeness.
]{
\includegraphics[width=0.47\textwidth]{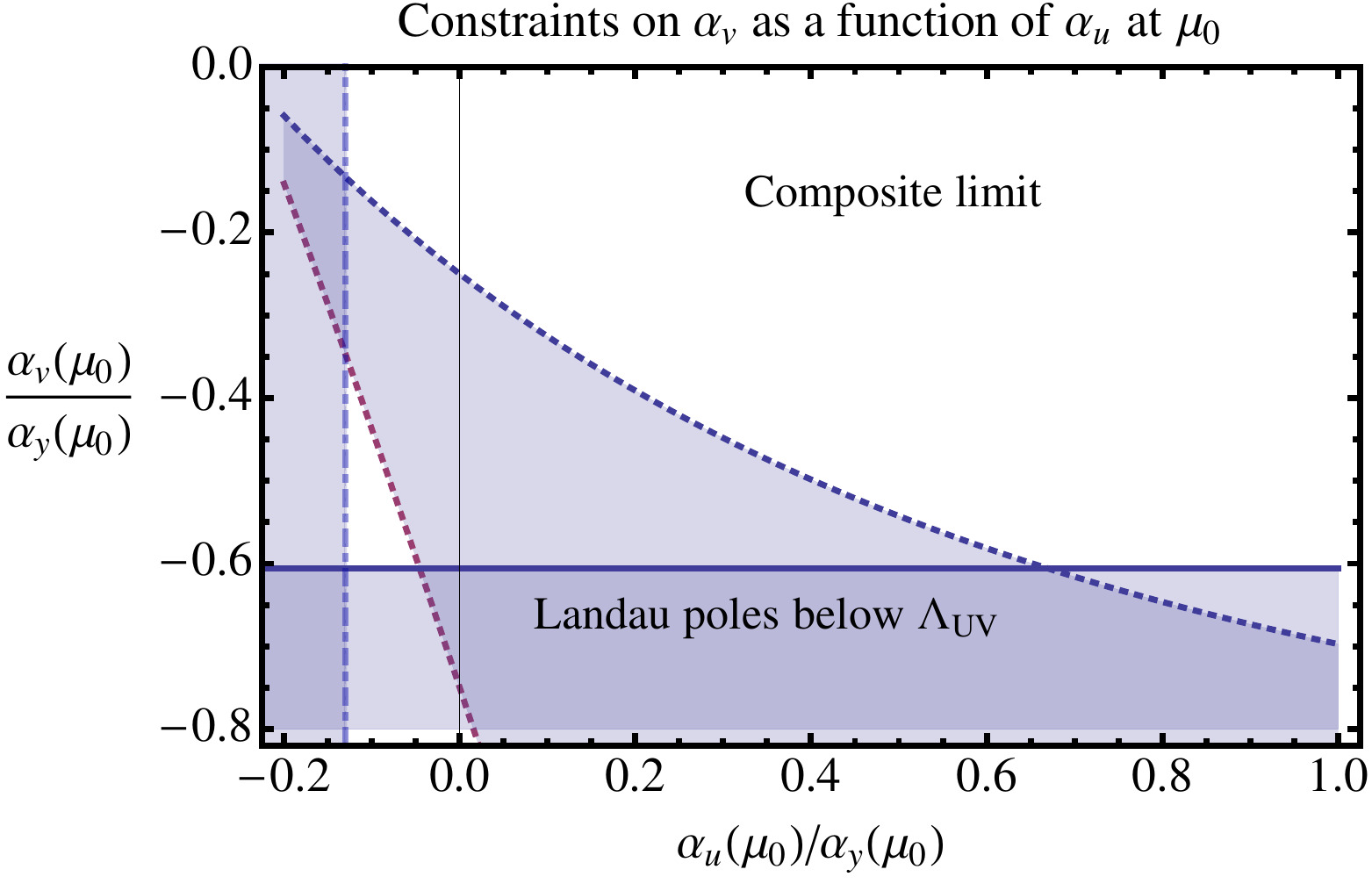}
\label{avvsaucon}}
\quad
\subfigure[\,
$x$-dependent bounds on compositeness for $\frac{\alpha_u(\mu_0)}{\alpha_y(\mu_0)} = -0.13$.]{
\includegraphics[width=0.47\textwidth]{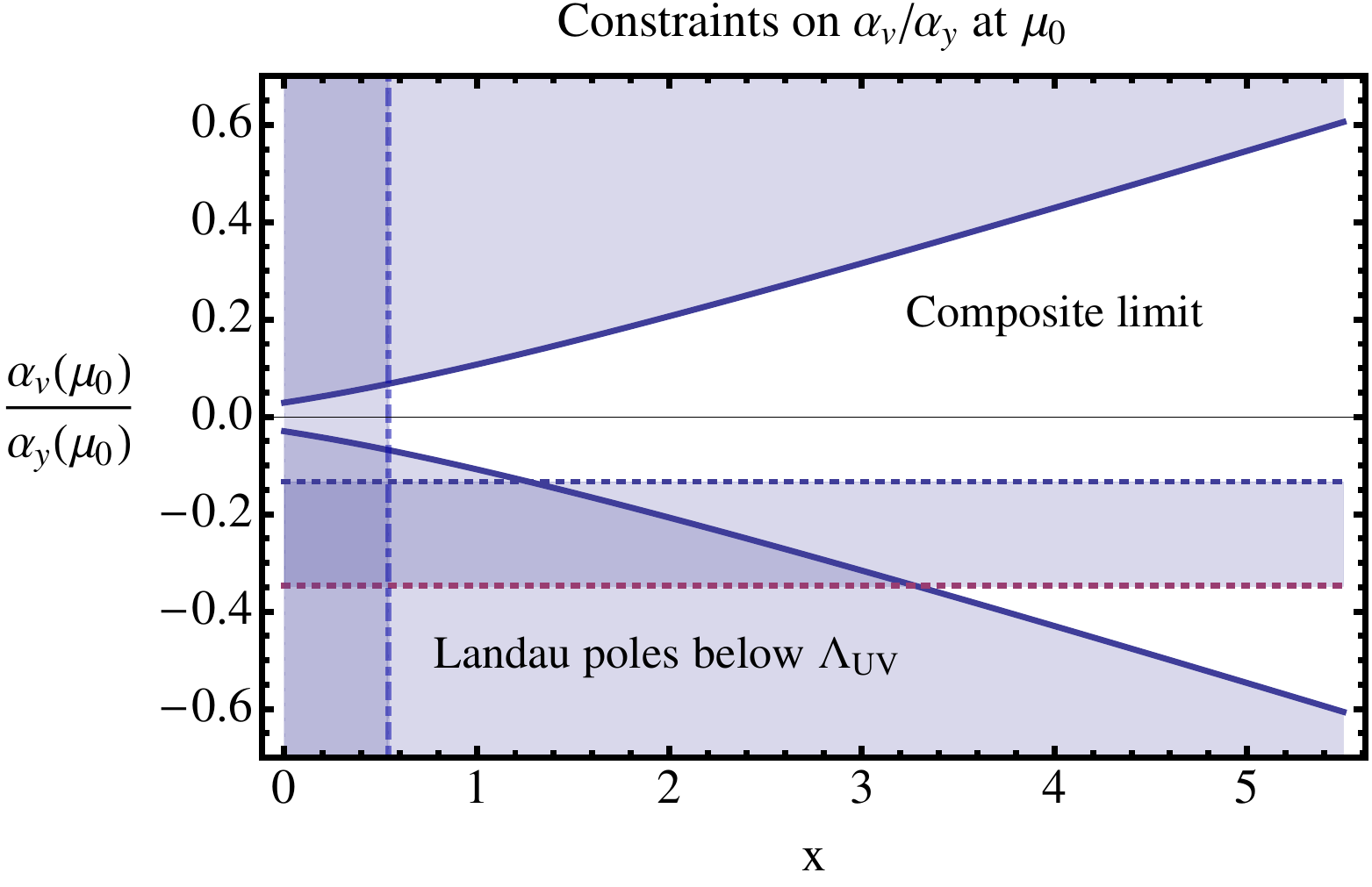}
\label{avconstraintsx}
}
\caption{Constraints on the initial value of the coupling $\alpha_v$ in terms of $\alpha_y$.
Fig.~\ref{avvsaucon} displays the $x$-independent constraints coming from 
Eq.~\eqref{avvsau1}, \eqref{avvsau2} (dots), and Eq.~\eqref{aucon3} (dotdashes), while the lowest $x$-dependent bound is also displayed (solid), which can be inferred from \ref{avconstraintsx}.
Fig.~\ref{avconstraintsx} displays the $x$-dependent strong coupling constraints
Eq.~\eqref{auavcon1} (solid), and Eq.~\eqref{last} (dotdashes), 
while the $x$-independent constraints Eq.~\eqref{avvsau1} and \eqref{avvsau2} (dots) are displayed for $\frac{\alpha_u(\mu_0)}{\alpha_y(\mu_0)} = -0.13$.
For larger values of $\frac{\alpha_u(\mu_0)}{\alpha_y(\mu_0)}$ the horizontal band moves downwards and closes the small window in the lower right corner for $\frac{\alpha_u(\mu_0)}{\alpha_y(\mu_0)} = -0.05$, as one can infer from \ref{avvsaucon}.}
\end{center}
\end{figure}

As mentioned, the picture for the other coupling, $\alpha_v$, is more involved, and the constraints depend on the values of $x$ and $\alpha_y$ as well as the ratio $\frac{\alpha_u}{\alpha_y}$. We start by examining the latter dependence, coming from Eq.~\eqref{avvsau1} and \eqref{avvsau2}, which is depicted in Fig~\ref{avvsaucon}.

The allowed regions for $\alpha_v(\mu_0)$ depend on the ratio $\alpha_u/\alpha_y$ in a nontrivial way, but notice that this dependence only constrains $\al v$ in the region of negative values, while leaving positive values for $\al v(\mu_0)$ unconstrained. 
 For values of ${\alpha_u}/{\alpha_y}$ larger than $ \sim 0.7$ the region excluded by Eq.~\eqref{avvsau1} is fully contained within the absolute lower bound coming from the strong coupling constraints Eq.~\eqref{auavcon1}, making the former constraint irrelevant. The region where Eq.~\eqref{avvsau1} and Eq.~\eqref{avvsau2} are most relevant, is the one where ${\alpha_u}/{\alpha_y}$ takes small values, the lowest value allowed from Eq.~\eqref{aucon3} being $\frac{\alpha_u(\mu_0)}{\alpha_y(\mu_0)} \simeq -0.13$.
In Fig.~\ref{avconstraintsx} we therefore display the strong coupling constraints Eq.~\eqref{auavcon1} (independent of $\al u/\al y$) alongside the constraints Eq.~\eqref{avvsau1} and \eqref{avvsau2}, evaluated at $\frac{\alpha_u(\mu_0)}{\alpha_y(\mu_0)} = -0.13$. For larger values of $\al u/\al y$, the horizontal band in Fig.~\ref{avconstraintsx} moves downwards and closes the small window of allowed parameter space in the lower right corner for $\frac{\alpha_u(\mu_0)}{\alpha_y(\mu_0)} = -0.05$, as one can infer from Fig.~\ref{avvsaucon}.

\begin{figure}[bt]
\begin{center}
\subfigure[\, 1-loop running couplings]{\includegraphics[width=0.47\textwidth]{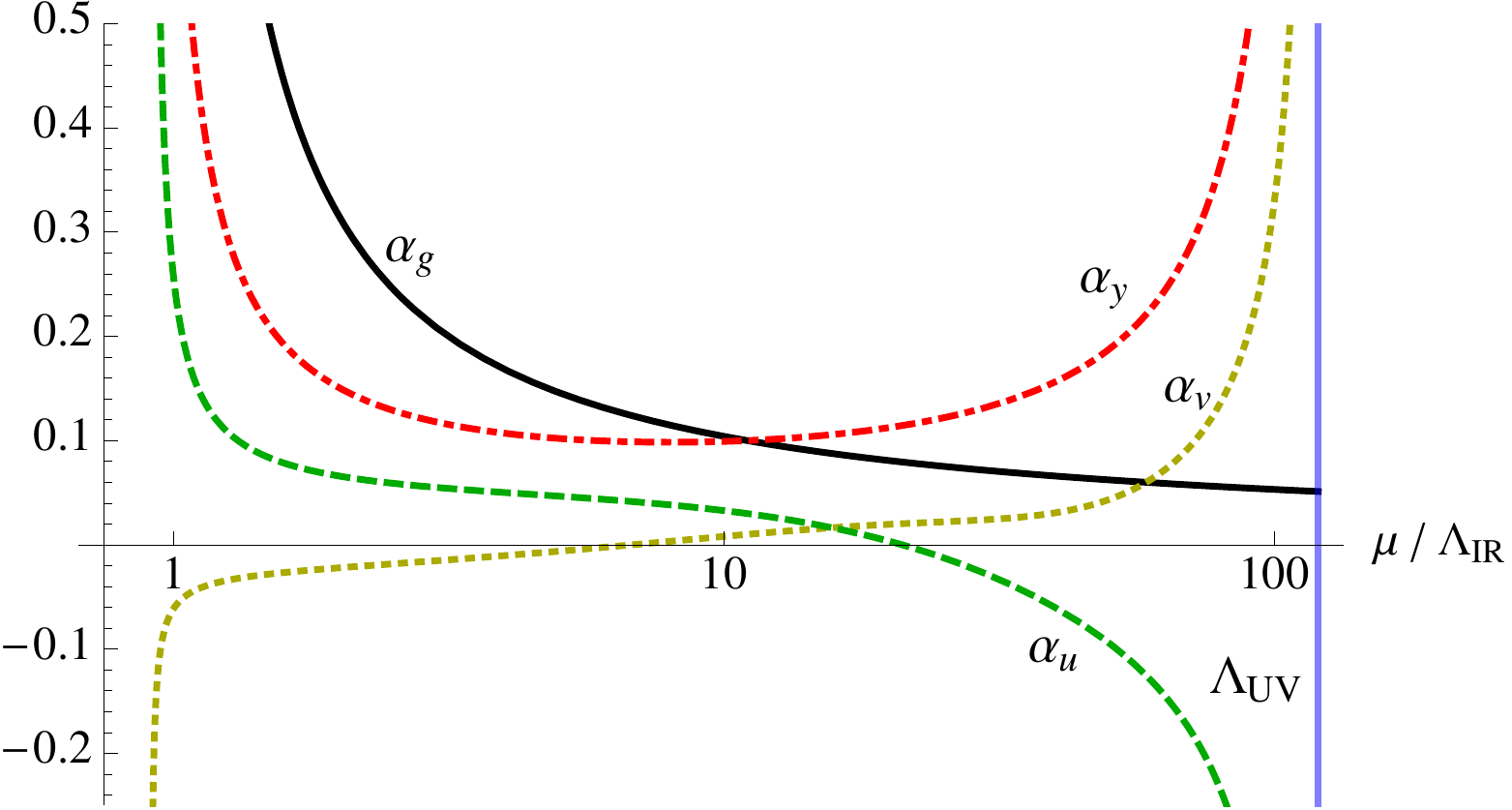}
\label{1run}
}
\quad
\subfigure[\, 1-loop running of coupling ratios]{\includegraphics[width=0.47\textwidth]{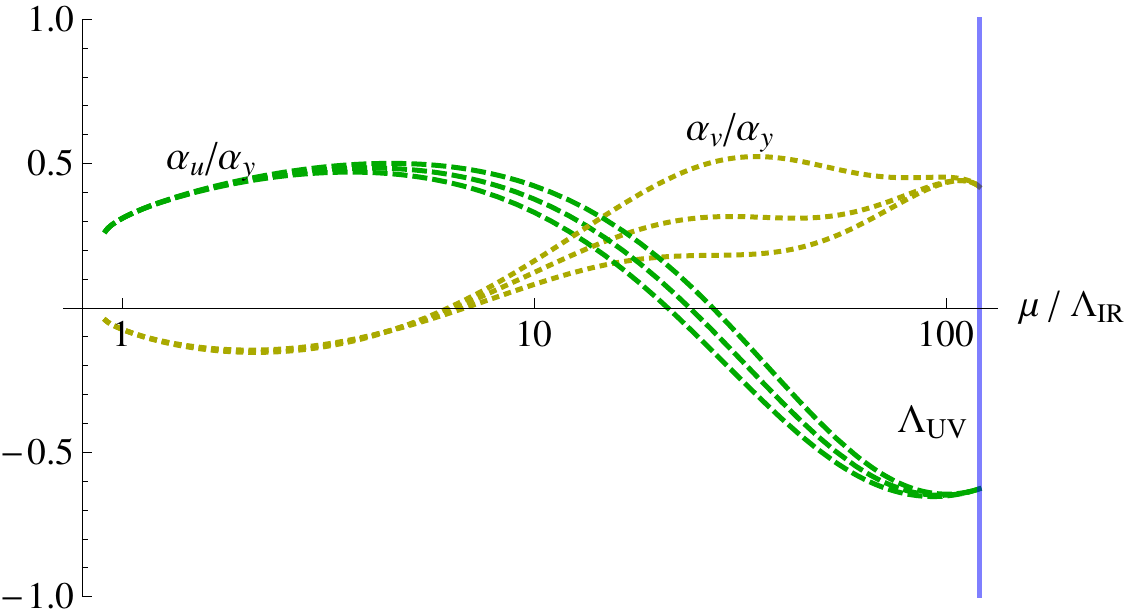}
\label{1ratio}
}
\caption{The RG evolution of the one loop system of gauge, Yukawa, and quartic couplings for a benchmark model, which respects the constraints for composite theories, where $x=2.5$ giving the smallest hierarchy between $\Lambda_{\rm UV}$ and $\Lambda_{\rm IR}$ (see Fig.~\ref{varyx}). 
Fig.~\ref{1run} shows the composite signature, where the divergence of the Yukawa and scalar couplings at $\Lambda_{\rm UV}$ is expected, and implies a possible composite interpretation of the theory.
Fig.~\ref{1ratio} shows that the ratios $\al v/\al y$ and $\al u/\al y$ are well-behaved in the entire region and run for different initial conditions to a unique fixed value at $\Lambda_{\rm UV}$, implying the possible composite theory to be of NJL-type.}
\label{111sketch}
\end{center}
\end{figure}

\begin{figure}[bt]
\begin{center}
\subfigure[\, NNLO running of the couplings]{
\includegraphics[width=0.47\textwidth]{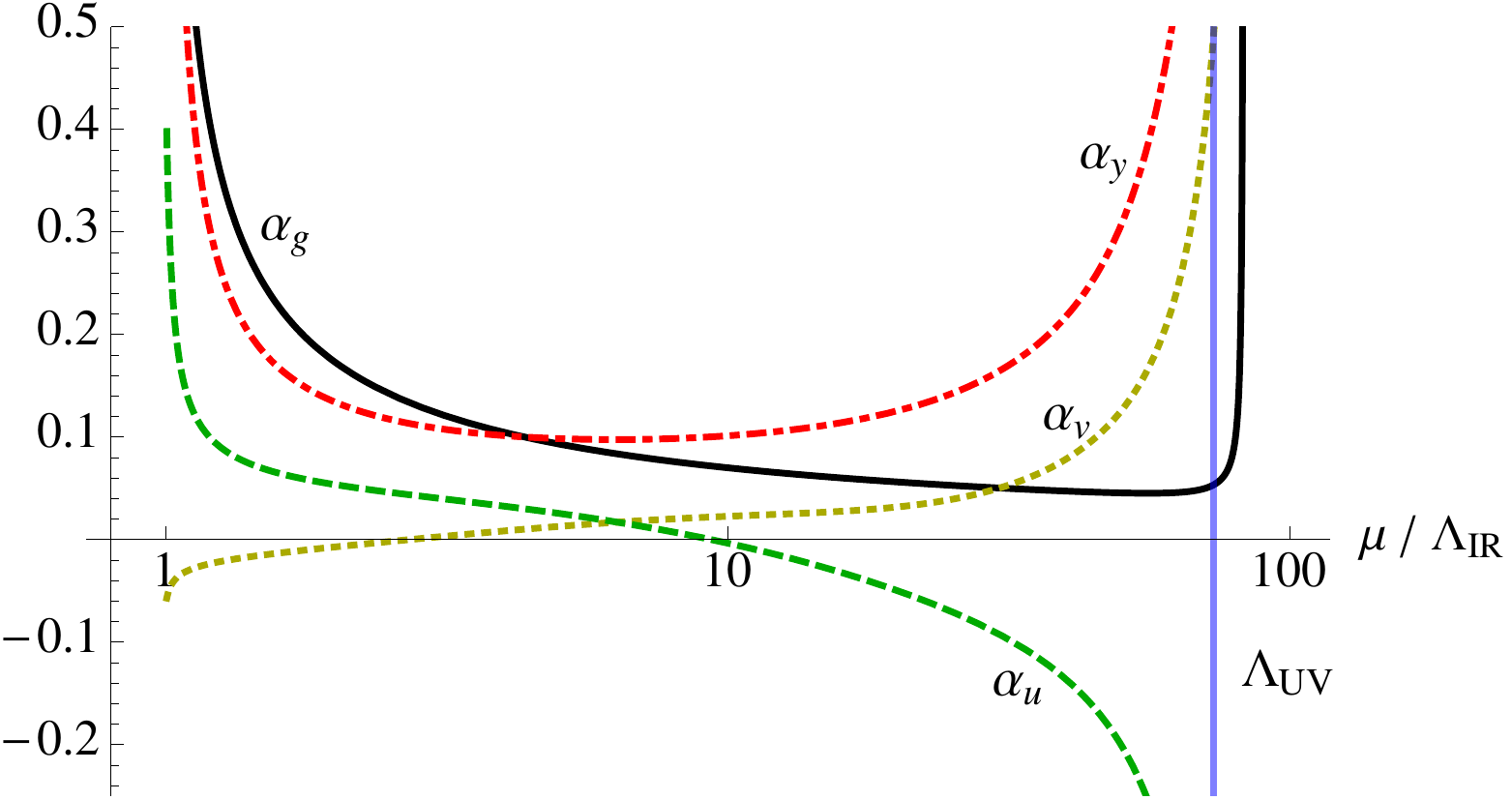}
 
\label{321sketch}
}
\quad
\subfigure[\, NNLO running of coupling ratios]{
\includegraphics[width=0.47\textwidth]{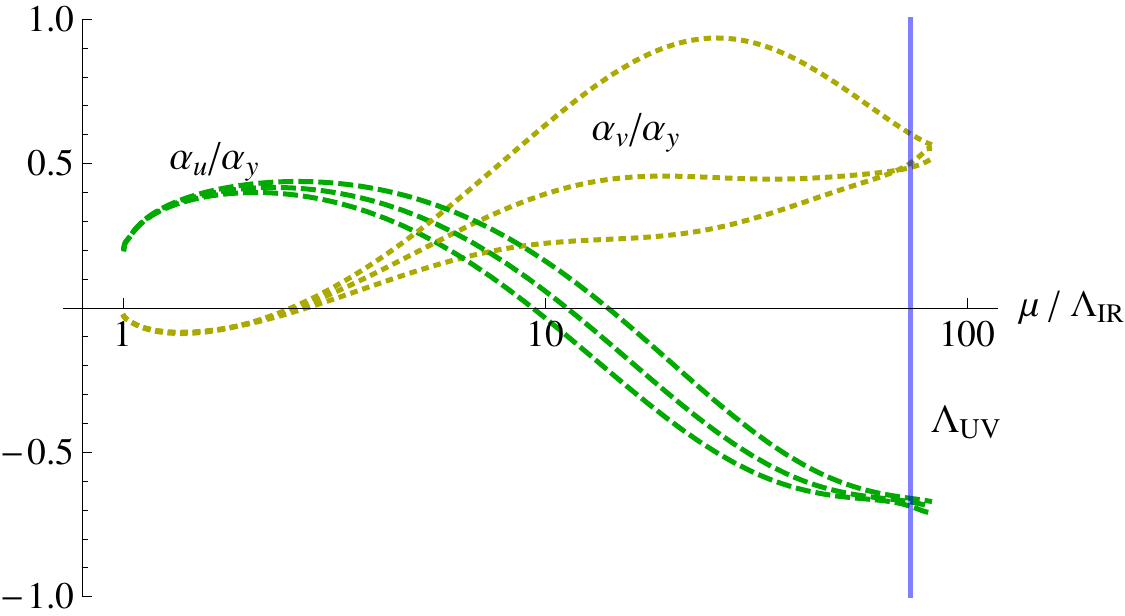}
\label{ratiosketch}
}
\caption{The evolution of the system of gauge, Yukawa, and quartic couplings for same values of parameters as in Fig~\ref{111sketch}. The running of the couplings is shown in \subref{321sketch}, while the evolution of some ratios of the couplings are shown in \subref{ratiosketch}. 
The conclusions are equivalent to that in Fig~\ref{111sketch}, however, he scale of IR divergence is now closer to the point where the couplings are initially defined, while the UV divergence is delayed by approximately the same amount. In addition, we also here see that the ratios between the quartic couplings and the Yukawa couplings stay well-defined, even when the couplings start to diverge, in accordance with the expected composite-like behavior of four-fermion interactions.
}
\vspace{-3mm}
\end{center}
\end{figure}

To test the validity of the approximations made in the calculations of the constraints above, we perform a full RG running of the coupled system of equations including the scalar couplings. As a benchmark model, we choose $x=2.5$ (giving the smallest hierarchy between $\Lambda_{\rm UV}$ and $\Lambda_{\rm IR}$, cf. Fig.~\ref{varyx}), and $\alpha_g(\mu_0)=\alpha_y(\mu_0)=0.1$, guaranteeing composite behavior in the gauge-Yukawa sector, as well as $\alpha_u (\mu_0)/\alpha_y(\mu_0) = 0.3$ and $\alpha_v (\mu_0)/\alpha_y(\mu_0) = 0.1$ to respect the constraints for the quartics. A numerical solution to the RG equations at one loop in all beta functions generates the running couplings shown in Fig.~\ref{111sketch}, where also the running of the ratios $\al u/\al y$ and $\al v/\al y$ is shown.
The result shows that the quartic couplings are well behaved between $\Lambda_{\rm IR}$ and $\Lambda_{\rm UV}$, where respectively the gauge and the Yukawa coupling poles are located.
The plot of the running of ratios demonstrates that they run to a unique constant at the composite scale, which signals that a possible composite UV completion is of four-fermion NJL-type.
Including the complete NNLO information in the RG equations, given at the beginning of this section, we find a very similar picture for the benchmark model, as shown in Fig.~\ref{321sketch}.
As predicted, we thus see that the initial conditions for the scalar couplings in this setup are not relevant for the UV behavior, in contrast to the situation for the simplest standard model extensions tailored for compositeness \cite{Krog:2015cna}

\begin{figure}[b]
\begin{center}
\includegraphics[width=0.5\textwidth]{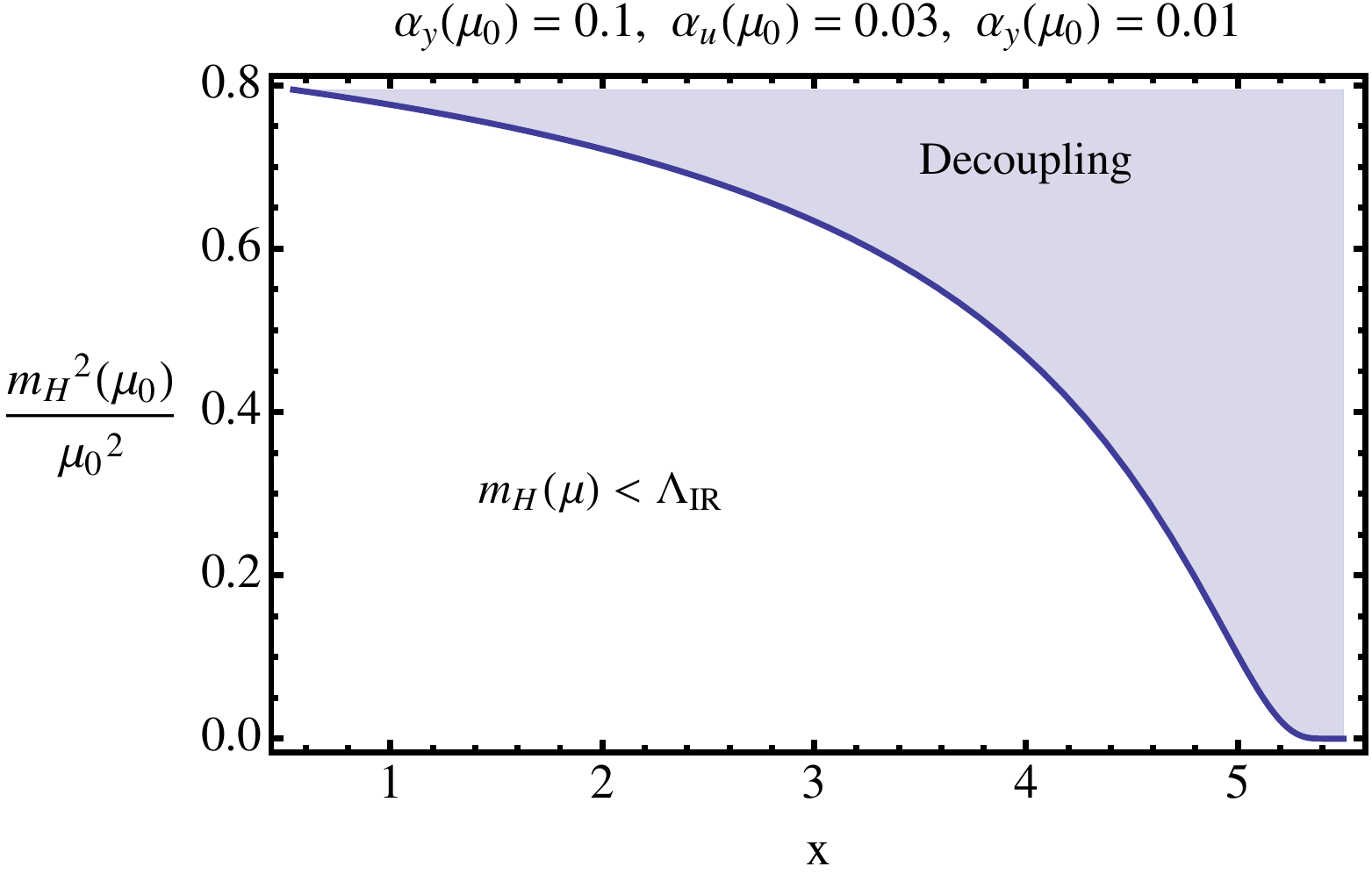}
\caption{Initial mass values $m_H^2(\mu_0)$ which in the white region respect the requirement Eq.~\eqref{masscon}, ensuring that the scalars do not decouple above the scale $\Lambda_{\rm IR}$.}
\label{massconfig}
\end{center}
\end{figure}

Next we consider the running of the mass and note that from its beta function the mass-squared parameter cannot change sign in perturbation theory. We further require the sign of $m_H^2$ to be positive to match the ultraviolet gNJL theory and to ensure stability of the scalar vacuum at the origin (for the Coleman-Weinberg instabilities concerning the case $m_H^2 = 0$ see~\cite{Bardeen:1993pj}).
The compositeness conditions tell us that the mass parameter must also diverge at the composite scale. At perturbative values, however, it must be ensured that $m_H(\mu) < \mu$ for every $\mu > \mu_0$, since otherwise the scalar fields would decouple at a scale $\mu^*$, where $m_H(\mu^*) = {\mu^*}$. In the perturbative regime, however, this can be easily achieved by choosing $m_H(\mu_0) < \mu_0$, since the growth in $m_H$ is logarithmic in $\mu/\mu_0$, and thus never exceeds $\mu$.
If we instead ask for the stronger constraint that the decoupling scale should be the strong IR scale of the previous sections, and not the $m_H$ scale, we need to impose the following constraint:
\ea{
m_H^2(\mu_0) < \Lambda_{\rm IR}^2 \left (\frac{\mu_0}{\Lambda_{\rm IR}} \right )^{\gamma_0} \ ,
}
where $\gamma_0$ is the one-loop coefficient of $\beta_{m_H^2}$ evaluated at $\mu_0$, i.e.
\ea{
\gamma_0 =  4 [ \alpha_y(\mu_0) +  \alpha_v(\mu_0) + 2 \al u(\mu_0) ]  \ .
}
Taking the IR scale to be the strong IR scale of the previous section, we get:
\ea{
m_H^2(\mu_0)
< \mu_0^2 \exp \left (- \frac{2- \gamma_0}{\beta_0 \al g(\mu_0)} \right ) \ .
\label{masscon}
}
This parameter choice ensures that the scale hierarchies computed in the previous section remain valid, when taking the scalar sector into account. One can imagine other possibilities that can lead to the generation of new intermediate scales with interesting phenomenological applications that we, however, do not consider here. We illustrate the requirement Eq.~\eqref{masscon} in Fig.~\ref{massconfig} for the benchmark parameters mentioned above while varying $x$.

Finally we must ensure stability of the potential. 
The scalar fields are well-defined in the regime \mbox{$\Lambda_{\rm IR} < H < \Lambda_{\rm UV}$}.
The scalar potential must for these values be positive to ensure the global minimum of the origin in field space. As shown in the appendix (cf. Eq.~\eqref{stabilitycond2}),
in the large $N_F$ limit the constraint ensuring stability of the potential reduces to
\ea{
m_H^2 > 0 \ .
}
Thus, if all the previous compositeness constraints are satisfied, the potential will automatically stay positive, in the entire domain of possible field values of $H$.

We have thus witnessed the emergence of a perturbative consistent picture of a subclass of gauge-Yukawa theories featuring Landau poles.  We argued that these theories suggest a composite picture because they are, in fact, gNJL theories.  We have furthermore shown that large scale hierarchies in these theories between the ultraviolet composite scale of the otherwise elementary scalar and the infrared scale leading to chiral symmetry breaking are a general feature. This result can be seen as the stepping stone towards realistic theories of SM fermion masses not at odds with flavour changing neutral currents constraints.

\section{Conclusions}
\label{Conclusions}
The discovery of the Higgs made imperative to explore the phase diagrams of nonsupersymmetric gauge-Yukawa theories and analyse their physical meaning. One interesting outcome has been the emergence of the first controllable four-dimensional example of a complete asymptotically safe theory \cite{Litim:2014uca,Litim:2015iea}. Here the  elementary scalars were needed by the dynamics to render the theory ultraviolet finite without the aid of any other super-imposed symmetry. 

On the other hand it is well known that the standard model is neither complete asymptotically free nor safe. That means that there are gauge-Yukawa theories that cannot be considered UV finite. In this work we therefore analysed the conditions that map gauge-Yukawa theories into time-honoured gauged four-fermion interactions.  Four fermion interactions emerge naturally when heavy degrees of freedom are integrated out. The set of extended compositeness conditions discussed here are non-perturbative in nature and permit us to investigate  theories of composite dynamics through gauge-Yukawa theories. 

We have shown that there are regimes along the RG flow in which the gauge-Yukawa description can be treated perturbatively.  We used as template an $SU(N_C)$ gauge theory featuring $N_F$ Dirac fermions transforming according to the fundamental representation of the gauge group. The fermions further interact with a gauge singlet complex $N_F\times N_F$ Higgs that ceases to be a propagating degree of freedom at the composite scale. 
We used the perturbative analysis to constrain the underlying gauge-Yukawa theory in order to enforce the compositeness conditions and showed that they can be nontrivially fulfilled.  

Within our example we argued that the theory leads to dynamical spontaneous symmetry breaking at an infrared dynamical scale $\Lambda_{\rm IR}$ and determined the ratio between the scale of the four-fermion interactions (the UV composite scale of the scalar $H$) and the infrared one as function of the external parameters of the theory, i.e. number of flavors and colors. We showed that these two scales are very well separated with a ratio that can easily be two orders of magnitude or much more. This would naturally allow to investigate, for example, the electroweak finite temperature phase transition within perturbation theory as recently summarised for a large number of gauge-Yukawa theories in~\cite{Sannino:2015wka}. 

Interestingly if one identifies the infrared composite scale with the electroweak scale  and the four-fermion interactions with the ones needed to give masses to the SM fermions, one discovers that in this type of theories the composite scalars can be light and the four-fermion interactions do not lead to flavour changing neutral currents.  Since, however, the main focus of our work is in elucidating the structure of the gauge-Yukawa theories we will leave the phenomenological analyses to another work. Nevertheless we cannot refrain from speculating that our results suggest the intriguing possibility that the standard model with its deceiving perturbative Higgs sector could hide, in plain sight, a composite theory. 

\acknowledgments 
{We thank Tuomas Karavirta for participation in the initial stages of the project.} 
This work is partially supported by the Danish National Research Foundation grant DNRF:90. 

%\clearpage

\appendix

\section{General analytic analysis of the compositeness conditions at one loop}

It is possible to study the compositeness conditions in a general perturbative gauge-Yukawa theory analytically, by analyzing the gauge-Yukawa-quartic system of beta functions, at one-loop in all couplings.
We will here consider the subspace of theories represented by the Lagrangian in Eqs.~\eqref{bosonized} and \eqref{TreePotential}.
The compositeness conditions on the couplings were given in Eq.~\eqref{Compositeness} and read:
\ea{
 \lim_{\mu\rightarrow\Lambda_{\rm UV}} \al y^{-1} = 0 \ ,
 \qquad \lim_{\mu\rightarrow\Lambda_{\rm UV}} \frac{\al u}{\al y^2} =\lim_{\mu\rightarrow\Lambda_{\rm UV}}\frac{\al v}{\al y^2}  = 0 \ ,
 \qquad \lim_{\mu\rightarrow\Lambda_{\rm UV}}\frac{y^2}{m_H^2} = \frac{G}{\Lambda_{\rm UV}^2} \, .
 \label{compcondA}
}

We imagine a situation where the theory considered is valid perturbatively around some energy scale $\mu_0$. We can then investigate, using the one-loop running of the couplings, whether the compositeness conditions will be satisfied at some higher scale~$\Lambda_{\rm UV}$. The one-loop beta-functions in the Veneziano limit of the theory will in general take the form:
\ea{
\beta_{\al g} &= \partial_t \al g = - \beta_0 \al g^2 \ , \\
\beta_{\al y} &= \partial_t \al y = \al y \left ( c_y \al y - c_g \al g\right ) \ , \\
\beta_{\al u} &= \partial_t \al u = \al u \left ( d_u \al u + d_y \al y \right) - d_{yy} \al y^2 \ , \\
\beta_{\al v} &= \partial_t \al v = \al v \left ( f_v \al v + d_y \al y + f_u \al u \right) + f_{uu} \al u^2  \, ,
}
where $t = \ln \mu/ \mu_0$ and all coefficients are positive definite in any gauge-theory, except for $\beta_0$. In infrared-free gauge theories $\beta_0 <0$. Here we do not consider such theories, as we require the gauge sector to be perturbatively well-defined at the composite scale, and thus require $\beta_0 >0$.
Notice that the coefficient $d_y$ is the same in $\beta_{\al u}$ and $\beta_{\al v}$ for any gauge theory. Also notice that $\beta_{\al u}$ is decoupled from $\al v$, which holds to all orders in the Veneziano limit.
The one-loop truncation of RG equations allows us to first solve the gauge sector, then the Yukawa sector, and finally the quartic sector sequentially.

\subsection{The gauge sector and the strong scale}
The solution of $\al g(t)$ is well known and reads:
\ea{
\frac{1}{\al g(t) } = \frac{1}{\al g(0) } + \beta_0 t \ .
\label{agt}
}
It has a strong confinement scale at the point, where the left hand side vanishes, which reads:
\ea{
t_s = \ln \frac{\Lambda_{\rm IR}}{\mu_0} = -\frac{1}{\beta_0\alpha_g(0)} \, ,
\label{strongscale}
}
where we defined $t_s$.

\subsection{The Yukawa sector and the composite scale}
The beta function $\beta_{\al g}$ can be used to reduce $\beta_{\al y}$ to an ordinary differential equation in terms of $R_{gy}= \al g/\al y$, which for $c_g \neq \beta_0$ reads:
\ea{
 \frac{ d R_{gy}}{d \ln \al g} = a ( R_{gy} - b ) \ ,
%c_y +\left ( \beta_0 - c_g\right ) R_{gy} \ ,
\label{Rgy}
}
where
\ea{ 
a = 1 - \frac{c_g}{\beta_0} \, , \quad b = \frac{c_y}{c_g - \beta_0} \ .
}
It follows that $a$ and $b$ have opposite signs and furthermore $a<1$.
The case $c_g = \beta_0$, where $b$ is not well-defined, will be considered in a moment.

It is easy to check that Eq.~\eqref{Rgy} has the solution:
\ea{
R_{gy}(t) = 
\left ( R_{gy}(0) - b \right )\left (\frac{\al g(t) }{\al g(0)} \right )^a + b \ .
%\left ( R_{gy}(0) - \frac{c_y}{\beta_0 + c_g} \right ) \exp\left (-\frac{\beta_0+c_g}{\beta_0} \right ) + \frac{c_y}{\beta_0 + c_g}
\label{rgy(t)}
}
The compositeness condition for $R_{gy}$ reads:
\ea{
R_{gy}(t_L)  = 0 \ , \quad \text{for} \quad 0<  t_L = \ln \frac{\Lambda_{\rm UV}}{\mu_0} < \infty \, ,
\label{rgycompcon}
}
where we defined $t_L$. Due to asymptotic freedom the last condition on $t_L$ can also be stated in terms of $\al g$:
\ea{
R_{gy}(t_L)  = 0 \ , \quad \text{for} \quad 1 > \frac{\al g(t_L) }{\al g(0)}  >0 \, .
}
It can then be seen that if $a>0$ (i.e. $c_g < \beta_0$), and thus $b <0$, the compositeness condition will always be satisfied.

On the other hand, if $a <0$ (i.e. $c_g > \beta_0$) we have to impose an extra condition,
since the composite scale in this case can be written as:
\ea{
\left (\frac{\al g(t_L) }{\al g(0)} \right )^{|a|} = 1-\frac{R_{gy}(0) }{b} \ .
}
The lower bound $t_L > 0$ (i.e. $\frac{\al g(t_L) }{\al g(0)} <1 $) is always satisfied, since $\frac{R_{gy}(0) }{b} >0$.
But the upper bound $t_L < \infty$ implies that $\frac{\al g(t_L) }{\al g(0)} >0$ and leads to a constraint on the parameter space:
\ea{
R_{gy}(0) < b   \ , \quad \text{(for $a<0$) .} 
}

Using the expression for $\al g(t)$ in Eq.~\eqref{agt} we can derive a general expression for the composite scale $\Lambda_{\rm UV}$, for any $a$ and $b$ satisfying the compositeness conditions:
\ea{
t_L= \ln \frac{\Lambda_{\rm UV}}{\mu_0} =\frac{1}{\beta_0\al g(0)}\left [\left(1-\frac{R_{gy}(0)}{b}\right)^{\frac{1}{a}}-1\right ] \ .
\label{compscale}
}

Finally, for the special case $a=0$, i.e. $c_g = \beta_0$, the RG equation for $R_{gy}$ reads:
\ea{
\frac{ d R_{gy}}{d \ln \al g}\Big |_{c_g = \beta_0} =  \frac{c_y}{\beta_0} \ .
}
From the general solution
\ea{
R_{gy}(t) = R_{gy}(0) + \frac{c_y}{\beta_0} \ln \frac{\al g(t)}{\al g(0)}  \ ,
}
it is readily seen that the compositeness condition parametrized by
\ea{
\al g(t_L) = \al g (0) \exp \left ( - \frac{\beta_0}{c_y} R_{gy}(0) \right ) < \al g(0) \ ,
}
is always satisfied, since the coefficients in the exponential are positive definite.
Furthermore the composite scale here reads:
\ea{
t_L= \ln \frac{\Lambda_{\rm UV}}{\mu_0} =\frac{1}{\beta_0\al g(0)}\left [\exp\left(\frac{\beta_0}{c_y}R_{gy}(0)\right)-1\right ] \ .
}
This is not in contradiction with Eq.~\eqref{compscale},
since it can be seen to be contained in that expression by noting that:
\ea{
\lim_{a\to 0} \left ( 1+ a \frac{R_{gy}(0)}{-ab} \right )^{\frac{1}{a}} = \exp\left ( \frac{R_{gy}(0)}{-ab} \right ) = \exp\left ( \frac{\beta_0}{c_y}R_{gy}(0)\right ) \, .
}

The hierarchy of scales between the composite and strong scales can now be computed:
\ea{
t_L-t_s = \text{ln} \frac{\Lambda_{\rm UV}}{\Lambda_{\rm IR}} = 
\frac{1}{\beta_0\al g(0)} \left(1-\frac{R_{gy}(0)}{b}\right)^{\frac{1}{a}}
 = \frac{1}{\beta_0\al g(0)}\left(1+R_{gy}(0)\frac{\beta_0-c_g}{c_y}\right)^{\frac{\beta_0}{\beta_0-c_g}} \ .
\label{generalrat}
}
As we noted before, the expression is regular for $(\beta_0 - c_g) \to 0$.

\subsubsection{The $SU(N)$ case}
Let us be specific and restrict to the case in Eq.~\eqref{venegauge2} discussed in the paper, where:
\ea{
\beta_0 = \frac{22-4 x}{3} \ , \quad c_y = 2(1+x)\ , \quad  c_g = 6 \ .
\label{specoef}
}
The parameters $a$ and $b$ then read:
\ea{
a = \frac{2(1-x)}{11-2x} \ , \qquad b = \frac{3(x+1)}{2(x-1)} \, .
}
For $x \leq 1$ we get that $a \geq 0$ and the compositeness conditions are always satisfied from the above analysis. For $x > 0 $ we get that $a < 0$ and $b >0$. The compositeness conditions are in this case only satisfied if furthermore $b > R_{gy}(0)$
or equivalently:
\ea{
\frac{\al y(0)}{\al g(0)}> \frac{2(x-1)}{3(x+1)} \ .
\label{compcond}
}
Since $\frac{\al y(0)}{\al g(0)} > 0$ is always true, this constraint holds
automatically for $x \leq 1$. Thus for any $x < 11/2$ (such that $\beta_0 >0$) we can uniquely impose the compositeness condition in Eq.~\eqref{compcond}.
Finally, the hierarchy of scales is given by:
 \ea{
 \ln \frac{\Lambda_{\rm UV}}{\Lambda_{\rm IR}}
%= \frac{\left(\frac{\alpha_g(\mu_0)}{\alpha_y(\mu_0)}\frac{b_0-c_1}{c_0}+1\right)^{\frac{1}{1-\frac{c_1}{b_0}}}}{b_0 \alpha_g(\mu_0)}\\
= \frac{3\left(1+\frac{\alpha_g(0)}{\alpha_y(0)}\frac{(1-x)}{3(1+x)}\right)^{2\frac{11-2x}{2(1-x)}}}{2(11-2x) \alpha_g(0)} \, .
}

\subsection{The quartic scalar sector}
From the compositeness conditions Eq.~\eqref{compcondA} it follows that
the quartic couplings may diverge only as fast as $\al y$ at the composite scale.
This specifically means that Landau poles in the quartic couplings entering before $t_L$, defined above, are not allowed. At the level of perturbation theory this is already implicit, since otherwise the above analysis would suffer from large corrections from the quartic couplings.

We consider, as before, the RG evolution of ratios. In particular, consider
\ea{
R_{yu} = \frac{\al y}{\al u} \ , \quad R_{yv} = \frac{\al y}{\al v}  \ .  
}
The RG equation for $R_{yu}$ can be written in terms of $R_{gy}$ as follows:
\ea{
\frac{d R_{yu}}{d \ln R_{gy}} = \frac{ d_u + c_g R_{gy} R_{yu} +(d_{y} - c_y) R_{yu} - d_{yy} R_{yu}^2}{c_y - (c_g - \beta_0) R_{gy} } \ .
\label{RyuRgy}
}
This equation is not well-defined at $c_y -(c_g-\beta_0)R_{gy}=0$, which is a problem we will get back to.
To investigate the compositeness conditions, however, we only need to understand the asymptotic behavior as $t \to t_L$, and since $R_{gy}(t_L) = 0$, and $c_y > 0$, the above equation is well-defined in limit $t \to t_L$. The asymptotic RG behavior is thus given by:
\ea{
\frac{d R_{yu}}{d \ln R_{gy}}\Big |_{t \to t_L} = \frac{ d_u}{c_y}  +\left (\frac{d_{y}}{c_y} - 1\right) R_{yu} - \frac{d_{yy}}{c_y} R_{yu}^2
= \rho_0 + \rho_1 R_{yu} - \rho_2 R_{yu}^2 \ ,
\label{RGRyu}
}
where to keep the notation light we introduced the coefficients $\rho_i$.
Defining some intermediary scale $t_* \lesssim t_L$, where the asymptotic solution is viable, we can parametrize this solution by:
\ea{
R_{yu}(t) \Big |_{t \approx t_L} = 
\frac{\rho_1 - \Delta_\rho \tanh \left ( K - \frac{\Delta_\rho}{2} \ln \frac{R_{gy}(t)}{R_{gy}(t_*)} \right )}{2 \rho_2} \ ,
}
where the discriminant $\Delta_\rho$ reads:
\ea{
\Delta_\rho = \sqrt{\rho_1^2 + 4 \rho_0 \rho_2} \ .
}
The integration constant $K$ is a number that has to be fixed by matching $R_{yu}(t_*)\Big |_{t \approx t_L}$ to the full solution given in terms of $R_{yu}(0)$ at the  scale $t_*$, and is for this analysis unimportant. 
The important result is that the solution exists and that the ratio of couplings $R_{yu}$ at the composite scale is fixed, since $R_{gy}(t_L) = 0$ and $\tanh( \infty) = 1$, and reads:
\ea{
R_{yu}(t_L) = \frac{\rho_1 - \Delta_\rho}{2 \rho_2}  \ ,
\label{RyutL}
}
which is also a fixed-point of the RG equation~\eqref{RGRyu}.
Notice that this value is negative, meaning that $\al u$ diverges to negative infinity as fast as $\al y$ diverges to positive infinity, while keeping their ratio constant.
This is potentially a problem for the stability of the potential near the composite scale. We shall comment on it after having considered the other quartic coupling $\al v$ as well.  
Let us comment on the region of validity of the above approximation. Since $R_{yu}(t_L)\neq 0$ for any parameter value, the asymptotic solution will be a good approximation as long as $R_{gy}(t) \ll 1$. This can be expressed in terms of the initial conditions:
\ea{
t_L > t_{\rm asymp.} \gg 
\frac{1}{\beta_0\al g(0)}\left [\left(\frac{R_{gy}(0)-b}{1-b}\right)^{\frac{1}{a}}-1\right ] \ ,
}
which for $a=0$ exponentiates to:
\ea{
t_L > t_{\rm asymp.} \gg 
\frac{1}{\beta_0\al g(0)}\left [\exp\left ( \frac{\beta_0}{c_y}\left(R_{gy}(0)-1\right )\right )-1\right ] \ .
}
Next we like to address the issue of divergence in Eq.~\eqref{RyuRgy}.
The potential problem is that if for some $t_s < t < t_L$ the denominator goes to zero, i.e.
$
R_{gy}(t) = \frac{c_y}{c_g-\beta_0} = b
$,
then the quartic coupling will diverge at $t$. 
If $\beta_0 > c_g$ then it is automatically never satisfied since $R_{gy}(t) >0$. Let us consider the case $c_g > \beta_0$, meaning that $a <0$ and $b>0$. From the general solution it is readily found that $R_{gy}(t)=b$ only occurs for $t = t_s$, which is consistent and not a problem.

Finally, as a last condition on $\al u$, we must ensure that it does not have Landau poles in the whole region $t_s < t < t_L$. We ensure this by negation: consider the case where $\al u$ does have a pole at a scale $t_s < t_u < t_L$.
Near this scale $\al u$ is much bigger than $\al y$, and to a good approximation the RG equation reads:
\ea{
\beta_{\al u} \approx d_u \al u^2 \ .
%\al u \left ( d_u \al u + d_y \al y (0) \right)
}
This is similar to the RG equation for $\al g$, and analogously its strong scale reads:
\ea{
t_u = \frac{1}{d_u \al u(0)}  \ .  
\label{tu}
}
To ensure perturbation theory to be valid in the region $t_s < t < t_L$ thus requires that $t_u < t_s$ or $t_u > t_L$. Formally this gives:
\ea{
d_u |\al u(0) | < \left \{ 
\begin{array}{llr} 
\beta_0 \al g(0) & \text{for } \al u(0)<0 & \qquad ( \Leftrightarrow t_u < t_s )\\
\beta_0 \al g(0)
\left [
\left(1-\frac{R_{gy}(0)}{b}\right)^{\frac{1}{a}}-1 \right ]^{-1} & \text{for } \al u(0) > 0 
 &\qquad ( \Leftrightarrow t_u > t_L )
\end{array} \right .
\label{aubounds}
}
One can also include the corrections to this, by including the term $\al y(t) \al u(t)$ in $\beta_{\al u}$ and setting $\al y (t) \approx \al y(0)$, which is a good approximation for intermediary scales. One then finds the strong scale of $\al u$ to be:
\ea{
t_u = \frac{1}{d_y \al y(0)} \log \left (1 + \frac{d_y}{d_u}\frac{\al y (0)}{\al u (0)} \right ) \ ,
}
which makes small corrections to the above bound on $\al u(0)$.
Finally, including all terms and assuming $\al y(t) \approx \al y(0)$, one can solve for $\al u(t)$ exactly. Defining $A = d_u$, $B = d_y \al y(0)$ and $C=- d_{yy} \al y(0)^2$, and the discriminant $D = \sqrt{B^2 - 4 A C}$, which is always real, since $C<0$, the solutions reads:
\ea{
\al u(t) \Big |_{t_{\rm intermed.}} = 
- \frac{B + D \tanh \left [ \frac{1}{2} D t -\tanh^{-1} \left (\frac{B + 2 A \al u(0)}{D} \right ) \right ] }{2 A}  \ .  
\label{auintermediate}
}
If the argument of $\tanh$ is real, there is never a Landau pole, since $\tanh \in [-1, 1]$ on the real domain. The argument can turn complex if $| B + 2 A \al u(0) | > D$, which potentially can lead to a Landau pole. Here one has to compute $t_u$ case by case and compare with $t_s$ and $t_L$. To avoid this, one can ensure that there is never a pole, by over-constraining the argument of $\tanh$ to always be real, i.e.:
\ea{
&| B + 2 A \al u(0) | < D \quad \Leftrightarrow \quad
- B - D < 2 A \al u(0) < - B + D
\label{aubound2}
\\
\Rightarrow &
-1-\sqrt{1+ \frac{4 d_u d_{yy}}{d_y^2}} 
 < 2 \frac{d_u \al u(0)}{ d_y \al y(0)} <
-1+\sqrt{1+ \frac{4 d_u d_{yy}}{d_y^2}} 
\ .
}

We now move on to the coupling $\al v$, through $R_{yv}$, as we did for $\al u$ above.
Its RG equation can be written as:
\ea{
\frac{d R_{yv}}{d \ln R_{gy}}=
\frac{\left(c_g
   R_{{gy}}-c_y+d_y\right) R_{{yu}}^2 R_{{yv}} +f_u R_{{yu}}R_{{yv}}+f_{{uu}} R_{{yv}}^2+f_v
   R_{{yu}}^2}{\left(R_{{gy}} \left(\beta
   _0-c_g\right)+c_y\right)R_{{yu}}^2
   }  \ .  
}
This is in general not a useful description, however, asymptotically the equation simplifies to:
\ea{
\frac{d R_{yv}}{d \ln R_{gy}}\Big |_{t\approx t_L}=
\frac{f_{uu}}{c_yR_{yu}^2(t_L)}R_{yv}^2 + \left (\rho_1 - \frac{f_u}{c_y| R_{yu}(t_L) | } \right )R_{yv}+ \frac{f_v}{c_y} 
= \eta_2 R_{yv}^2 + \eta_1 R_{yv} + \eta_0 \ ,
}
where we defined the coefficients $\eta_i$. Note that $\eta_0>0$ and $\eta_2>0$, while $\eta_1$ can take any real value, in general.
The general solution reads:
\ea{
R_{yv}(t)\Big|_{t \approx t_L} =-\frac{\eta_1 + \Delta_\eta \tanh \left ( \frac{1}{2} \Delta_\eta \ln \frac{R_{gy}(t)}{R_{gy}(t_*)} + K_v \right )}{2 \eta_2} \ ,
}
where $t_*$ is defined as before and
\ea{
\Delta_\eta = \sqrt{\eta_1^2 - 4 \eta_0 \eta_2} > 0 \ .
\label{Dconstraint}
}
The positivity constraint on this expression is a requirement we have to impose to satisfy the compositeness conditions; for imaginary $\Delta_\eta$ the above expression switches from $\tanh$ to $\tan$, and leads to Landau poles in $\al v$ before the composite scale.
 This is therefore a constraint on the possible theory space of gauge-Yukawa theories we are considering.
Furthermore we get that:
\ea{
R_{yv}(t_L) = - \frac{\eta_1 - \Delta_\eta}{2 \eta_2} \ .
\label{RyvtL}
}

Finally, we repeat the exercise of removing possible parameter region that violates perturbation theory in the region $t_s < t < t_L$ by considering the strong scale of $\al v$. To a first approximation it is simply:
\ea{
t_ v = \frac{1}{f_v \al v(0)} \ ,
\label{tv}
}
which leads to the equivalent bounds as in Eq.~\eqref{aubounds}.
Perturbation theory is ensured if $t_v < t_s$ for $\al v(0) < 0$ and $t_v > t_L$ for $\al v(0) > 0$. Finally, we can again solve the full differential equation by assuming that at intermediate scales $\al y(t) \approx \al y(0)$ and $\al u (t) \approx \al u(0)$, which are good approximations in the composite phase space. Defining this time $A=f_v$, $B= d_y \al y(0) + f_u \al u (0)$ and $C = f_{uu} \al u(0)^2$ and the discriminant $D= \sqrt{B^2-4 A C}$, the solution is given by the same expression as for $\al u$ in Eq.~\eqref{auintermediate}. However, note that this time $C>0$ and thus the discriminant can turn complex, i.e. for $B^2 < 4 A C$. Considering this case, the expression is rewritten in terms of $\tan$:
\ea{
\al v (t) \Big |_{t_{\rm intermed.}} =
- \frac{B - (i D) \tan \left [ \frac{1}{2} (iD) t +\tan^{-1} \left (\frac{B + 2 A \al v(0)}{(iD)} \right ) \right ] }{2 A}  \ .  
\label{avintermediate}
}
In this case there are many poles, since $\tan(\pi/2 + n\pi) = \pm \infty$ for all integer $n$. The scales at which these occur is given by:
\ea{
t_v = \frac{\pi - 2 \tan^{-1} \left (\frac{B + 2 A \al v(0)}{(iD)} \right ) + n \cdot 2 \pi}{(iD)} \ .
\label{tv3}
}
This leads to the extra constraint, i.e. the smallest negativ $t_v$ has to be less than $t_s$ and the smallest positive $t_v$ has to be bigger than $t_L$. This constraint is relevant whenever $4 f_v f_{uu} > \left ( f_u + d_y \frac{\al y(0)}{\al u(0)}\right )^2$, while in the opposite case one should consider a constraint equivalent to Eq.~\eqref{aubound2}.

\subsubsection{The $SU(N)$ case}
Let us apply the above analysis to the case considered in the paper. The beta function coefficients read:
\ea{
&d_u = 8 \, , \ d_y = 4 \, , \ d_{yy} = 2 x \, , \\
&f_v = 4 \, , \ f_u = 16 \, , \ f_{uu} = 12  \ .  
}
From these we derive the relevant parameters:
\ea{
&
\rho_0 = \frac{4}{1+x} \, , \ \rho_1 = \frac{1 - x}{1+x} \, , \ \rho_2 = \frac{x}{1+x} 
%\\
%&
\, , \
\Delta_\rho = \frac{\sqrt{(1-x)^2 + 16 x}}{1+x} \, 
, 
\\[2mm]
&
\eta_0 = \frac{2}{1+x} \, , \ \eta_1 = - \frac{\sqrt{(1-x)^2 + 16 x}}{1+x}
\, , \ 
\eta_2 = \frac{24 x^2}{(1+x)(x-1+ \sqrt{(1-x)^2 + 16 x})^2}  \ .  
}
Notice that $\eta_1 <0$ for any $x$. The expression for $\Delta_\eta$ takes a lengthy expression, but its constraint Eq.~\eqref{Dconstraint} leads to:
\ea{
x > - 4 + 3 \sqrt{3} - \sqrt{6(7-4\sqrt{3})} \approx 0.54 \ .
}

For a given $x$ it is always possible to find initial parameter values for $\al g$, $\al y$, $\al u$ and $\al v$ such that the constraints Eqs.~\eqref{compcond}, \eqref{aubounds}, \eqref{aubound2}, \eqref{Dconstraint}, and the ones related to Eq.~\eqref{tv}-\eqref{tv3} are satisfied. We consider the details in the paper.

We have furthermore found that the ratio of quartic couplings over $\al y$ are completely fixed at the composite scale, independent of initial conditions, and given by Eq.~\eqref{RyutL} and \eqref{RyvtL}.

\subsection{Running mass and stability of the potential}

The RG equation describing the running of the scalar mass term is given by
\begin{equation}
\label{MRGE}
 \beta_{m_H^2} = \partial_t m_H^2 = m_H^2 (h_y \alpha_y + h_v \alpha_v + h_u \alpha_u )  \, ,
\end{equation}
where for $SU(N_C)$ the parameters read in the Veneziano limit: $h_y = 4 \, , \ h_v = 4 \, , \ h_u = 8$.
This expression shows that, where it is valid, the scalar mass term can not change sign, since its beta function is proportional to the squared mass itself. The initial condition $m_H^2 > 0$, should then ensure that spontaneous symmetry breaking will not occur in the range where the above expression may be applied.

The further condition $m_H^2(\mu)<\mu^2$ should also be satisfied to make sure that no scalar states decouple at energies higher than the strong scale. However, this constraint is not related to compositeness and can be relaxed to instead read $m_H^2(\mu) < \mu_0$ such that the composite nature of the theory, which is probed for $\mu > \mu_0$ stays intact, while the IR physics defined by $\mu < \mu_0$ may have different phases. In our analysis we have constraint the IR phase to be dominated by strong gauge interactions.

%The beta function of the mass can also be written in terms of the dimensionless parameter $\tilde{m}^2 = m_H^2/\mu^2$, which reads instead:
%\ea{
%\beta_{\tilde{m}^2} = \tilde{m}^2 \left (
%h_y \alpha_y + h_v \alpha_v + h_u \alpha_u - 2 \right ) \ .
%}

Proceeding to study the stability conditions on the potential, we first note that the scalar fields are well-defined for field values $m_H(\mu) < H < \Lambda_{\rm UV}$, where as argued above $m_H(\mu) < \mu_0$.
For a positive mass-term it is clear that the potential has a minimum at the origin, $\langle H \rangle = 0$, which preserves the $U(N_F) \times U(N_F)$ symmetry of the classical theory. To ensure consistency of our analysis, we must make sure that this symmetry is obeyed for large field values as well and at every scale in the region $\mu_0 < \mu < \Lambda_{\rm UV}$. It is enough to study the diagonal field $H = {\rm diag(h_1, \ldots, h_{N_F})}$ since this can be rotated into any other $H$ by a $U(N_F) \times U(N_F)$ transformations. In terms of $h_i$ the potential reads:
\ea{
V = m_H^2 \sum_{i=1}^{N_F} h_i^2 + u \sum_{i=1}^{N_F} h_i^4 + v \left (\sum_{i=1}^{N_F} h_i^2\right )^2  \ .  
}
We consider the general case where $u$ and $v$ can take both positive and negative values. 
As argued before if all $h_i$ are small (i.e. $h_i \ll m_H$) then one sees that the minimum is at the origin, since $m_H >0$.
Let us now consider what happens for large values of some of the fields $h_i$, in particular take for $i= 1,\ldots, n$ the fields $h_i \to \Lambda_{\rm UV}$, while for $i=n+1,\ldots, N_F$ keep $h_i \ll m_H$.
Then the potential is dominated by the large fields and reads approximately:
\ea{
V \approx m_H^2 (n \Lambda_{\rm UV}^2) + u (n\Lambda_{\rm UV}^4) + v(n^2 \Lambda_{\rm UV}^4)  \ .  
}
Positivity of this potential requires:
\ea{
\frac{m_H^2}{n \Lambda_{\rm UV}^2} + \frac{u}{n} + v \geq 0  \ .  
\label{stabilitycond}
}
In term of the rescaled couplings from Eq.~\eqref{couplings}, this becomes
\ea{
\frac{m_H^2N_F^2}{n(4\pi\Lambda_{\rm UV})^2} + \frac{\alpha_u N_F}{n} + \alpha_v \geq 0   \quad \stackrel{N_F \to \infty}{\Longrightarrow} \quad
m_H^2 > 0 
\ .  
\label{stabilitycond2}
}
Thus, in the large $N_F$ limit already assumed constraint $m_H^2 > 0$ ensures the potential to stay positive in the entire region of field values.

For completeness, let us discuss the finite $N_F$ case, and thus consider the unrescaled couplings. If $u$ is negative, then the strongest constraint comes from $n=1$, yielding the constraint:
\ea{
\frac{m_H^2 }{\Lambda_{\rm UV}^2} + v \geq - u \quad \text{for } u < 0 \ .
}
If $v$ is negative, the strongest constraint comes from $n = N_F$, thus:
\ea{
\frac{m_H^2 }{\Lambda_{\rm UV}^2} + u \geq - v \  N_F \quad \text{for } v < 0 \ .
}
If both $u$ and $v$ are negative, one has to maximize the function $u + n v$ for $n$, and ensure that the general constraint Eq.~\eqref{stabilitycond} is satisfied.

\end{document}